\documentclass[letterpaper]{article} 
\usepackage{aaai23}  
\usepackage{times}  
\usepackage{helvet}  
\usepackage{courier}  
\usepackage[hyphens]{url}  
\usepackage{graphicx} 
\usepackage{natbib}  
\usepackage{caption} 
\usepackage{algorithm}
\usepackage{algorithmic}

\usepackage{newfloat}
\usepackage{listings}
\DeclareCaptionStyle{ruled}{labelfont=normalfont,labelsep=colon,strut=off} 
\floatstyle{ruled}
\newfloat{listing}{tb}{lst}{}
\floatname{listing}{Listing}
\usepackage{amsmath}
\usepackage{amsfonts}
\usepackage{booktabs}
\usepackage{diagbox}
\usepackage{makecell}
\usepackage{multirow}
\usepackage{subfigure}
\title{Deep Spiking Neural Networks with High Representation Similarity Model Visual Pathways of Macaque and Mouse}
\author {
Liwei Huang\textsuperscript{\rm 1,\rm 2},
Zhengyu Ma\textsuperscript{\rm 2}\footnotemark[1],
Liutao Yu\textsuperscript{\rm 2},
Huihui Zhou\textsuperscript{\rm 2},
Yonghong Tian\textsuperscript{\rm 1,\rm 2}\thanks{Corresponding author.}
}
\affiliations {
\textsuperscript{\rm 1}National Engineering Research Center of Visual Technology, School of Computer Science, Peking University, China\\
\textsuperscript{\rm 2}Department of Networked Intelligence, Peng Cheng Laboratory, China\\
huanglw20@stu.pku.edu.cn, \{mazhy, yult, zhouhh\}@pcl.ac.cn, yhtian@pku.edu.cn
}

\begin{document}

\maketitle

\begin{abstract}
	Deep artificial neural networks (ANNs) play a major role in modeling the visual pathways of primate and rodent. However, they highly simplify the computational properties of neurons compared to their biological counterparts. Instead, Spiking Neural Networks (SNNs) are more biologically plausible models since spiking neurons encode information with time sequences of spikes, just like biological neurons do. However, there is a lack of studies on visual pathways with deep SNNs models. In this study, we model the visual cortex with deep SNNs for the first time, and also with a wide range of state-of-the-art deep CNNs and ViTs for comparison. Using three similarity metrics, we conduct neural representation similarity experiments on three neural datasets collected from two species under three types of stimuli. Based on extensive similarity analyses, we further investigate the functional hierarchy and mechanisms across species. Almost all similarity scores of SNNs are higher than their counterparts of CNNs with an average of $6.6\%$. Depths of the layers with the highest similarity scores exhibit little differences across mouse cortical regions, but vary significantly across macaque regions, suggesting that the visual processing structure of mice is more regionally homogeneous than that of macaques. Besides, the multi-branch structures observed in some top mouse brain-like neural networks provide computational evidence of parallel processing streams in mice, and the different performance in fitting macaque neural representations under different stimuli exhibits the functional specialization of information processing in macaques. Taken together, our study demonstrates that SNNs could serve as promising candidates to better model and explain the functional hierarchy and mechanisms of the visual system.
\end{abstract}

\section{Introduction}
\label{intro}

Originally, the prototype of deep neural networks is inspired by the biological vision system \cite{hubel1959receptive, hubel1962receptive}. To date, deep neural networks not only occupy an unassailable position in the field of computer vision \cite{lecun2015deep}, but also become better models of the biological visual cortex compared to traditional models in the neuroscience community \cite{khaligh2014deep, yamins2014performance, yamins2016using}. They have been successful at predicting the neural responses in primate visual cortex, matching the hierarchy of ventral visual stream \cite{gucclu2015deep, kubilius2019brain, nayebi2018task, kietzmann2019recurrence}, and even controlling neural activity \cite{bashivan2019neural, ponce2019evolving}. Moreover, as training paradigms of mice \cite{zoccolan2009rodent} and techniques for collecting neural activity \cite{de2020large} have been greatly improved, there is a strong interest in exploring mouse visual cortex. Deep neural networks also play an important role in revealing the functional mechanisms and structures of mouse visual cortex \cite{shi2019comparison, cadena2019well, nayebi2022mouse, bakhtiari2021functional, conwell2021neural}.

Compared to biological networks, Artificial Neural Networks discard the complexity of neurons \cite{pham2008control}. Spiking Neural Networks, incorporating the concept of time and spikes, are more biologically plausible models \cite{maass1997networks}. To be more specific, because of their capabilities of encoding information with spikes, capturing the dynamics of biological neurons, and extracting spatio-temporal features, deep SNNs are highly possible to yield brain-like representations \cite{hodgkin1952quantitative, gerstner2002spiking, izhikevich2004model, brette2007simulation, kasabov2013dynamic}. However, deep SNNs have not been employed to model visual cortex due to the immaturity of training algorithms. Recently, a state-of-the-art directly trained deep SNN \cite{fang2021deep}, makes it possible to use deep SNNs as visual cortex models.

\begin{figure*}[t]
	\centering
	\includegraphics[width=0.99\textwidth]{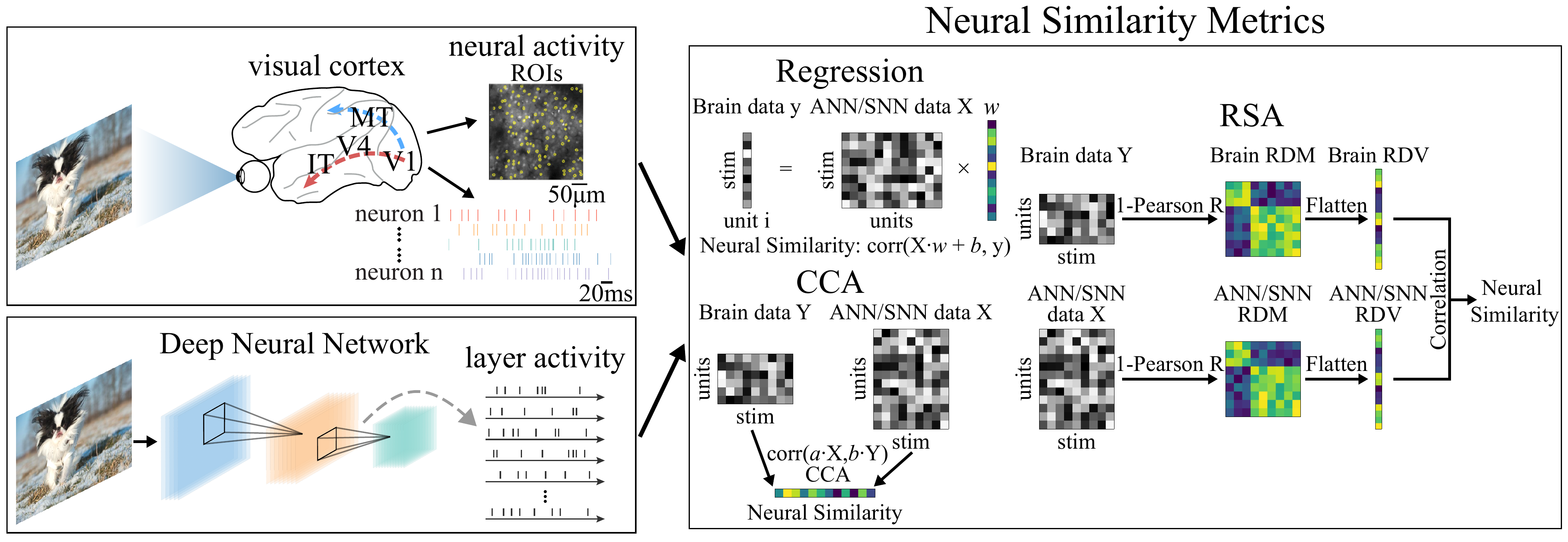}
	\caption{To conduct neural representation similarity experiments, we apply three similarity metrics to a layer-by-layer comparison between the responses of models and the neural activities of visual cortex.}
	\label{Fig.flowsheet}
\end{figure*}

\textbf{Contributions.} In this work, we conduct large-scale neural representation similarity experiments on SNNs and other high-performing deep neural networks to study the brain’s visual processing mechanisms, with three datasets and three similarity metrics (Figure \ref{Fig.flowsheet}). Specifically, to the best of our knowledge, we are the first to use deep SNNs to fit complex biological neural representations and explore the biological visual cortex. We summarize our main contributions in four points as follows.

\begin{itemize}
	\item We find that SNNs outperform their counterparts of CNNs with the same depth and almost the same architectures in almost all experiments. In addition, even with very different depths and architectures, SNNs can achieve top performance in most conditions.
	\item By making a more direct comparison between macaques and mice for the first time, we reveal the differences in the visual pathways across the two species in terms of the homogeneity of visual regions and the increases of receptive field sizes across cortical visual pathways, which is consistent with previous physiological work. 
	\item The multi-branch structures in neural networks benefit neural representation similarity to mouse visual cortex, providing computational evidence that parallel information processing streams are widespread between cortical regions in the mouse visual system.
	\item Comparing the results of two macaque neural datasets under different stimuli, we reveal that the macaque vision system may have functional specialization for processing human faces and other natural scenes.
\end{itemize}

Altogether, as the first work to apply deep SNNs to fit neural representations, we shed light on visual processing mechanisms in both macaques and mice, demonstrating the potential of SNNs as a novel and powerful tool for research on the visual system. Our codes and appendix are available at \textit{https://github.com/Grasshlw/SNN-Neural-Similarity}.

\section{Related Work}
\label{related-work}

There are plenty of computational models of macaque and mouse visual systems for exploring the visual processing mechanisms recently. We summarize some of the outstanding work in the following.

\textbf{The network models of macaque visual system.} In the early days, studies basically used simple feedforward neural networks as the models of the macaque visual system \cite{khaligh2014deep, yamins2014performance, yamins2016using}. Recently, some bio-inspired or more complex models achieved better performance in fitting the neural representations of macaque visual cortex \cite{kubilius2019brain, dapello2020simulating, zhuang2021unsupervised, higgins2021unsupervised}. \cite{kubilius2019brain} proposed a brain-like shallow CNN with recurrent connections to better match the macaque ventral visual stream. By mimicking the primary stage of the primate visual system, VOneNets \cite{dapello2020simulating} performed more robustly in image recognition while better simulating macaque V1. Moreover, the representations learned by unsupervised neural networks \cite{zhuang2021unsupervised, higgins2021unsupervised} also effectively matched the neural activity of macaque ventral visual stream. Although the above work developed many bio-inspired structures, the networks are still traditional ANNs in nature. Our work introduces deep SNNs for the first time to explore the visual processing mechanisms of macaque visual system.

\textbf{The network models of mouse visual system.} Large-scale mouse neural dataset provided an experimental basis for model studies of mouse visual system \cite{de2020large, siegle2021survey}. \cite{shi2019comparison} conducted comparisons between the representations of mouse visual cortex and the VGG16 trained on the ImageNet dataset. In \cite{bakhtiari2021functional}, they developed a single neural network to model both the dorsal and ventral pathways with showing the functional specializations. What's more, a large survey of advanced deep networks \cite{conwell2021neural} revealed some hierarchy and functional properties of mice. Similar to the studies of macaque visual system, deep SNNs have never been used to model the mouse visual system. In this work, we not only use SNNs as one of the candidates to fit the representations of mouse visual cortex, but also conduct direct comparisons between macaques and mice to further investigate the functional hierarchy and mechanisms of the two species.

\section{Methods}
\label{methods}

\subsection{Neural Datasets}
\label{methods.dataset}

Our work is conducted with three neural datasets. These datasets are recorded from two species under three types of stimuli. More specifically, there are neural responses of mouse visual cortex to natural scene stimuli, and responses of macaque visual cortex to face image and synthetic image stimuli.

\textbf{Allen Brain mouse dataset.} It is part of the Allen Brain Observatory Visual Coding dataset \cite{siegle2021survey} collected using Neuropixel probes from 6 regions simultaneously in mouse visual cortex. Compared to two-photon calcium imaging, Neuropixel probes simultaneously record the spikes across many cortical regions with high temporal resolution. In these experiments, mice are presented with 118 250-ms natural scene stimuli in random orders for 50 times. Hundreds to thousands of neurons are recorded for each brain region. To get the stable neurons, we first concatenate the neural responses (average number of spikes in 10-ms bins across time) under 118 images for each neuron, and then preserve the neurons whose split-half reliability across 50 trials reaches at least 0.8.

\textbf{Macaque-Face dataset.} This dataset \cite{chang2021explaining} is composed of neural responses of 159 neurons in the macaque anterior medial (AM) face patch under 2,100 real face stimuli, recorded with Tungsten electrodes. For this dataset, we compute the average number of spikes in a time window of 50-350ms after stimulus onset and exclude eleven neurons with noisy responses by assessing the neurons' noise ceiling. The details of the preprocessing procedure are the same as \cite{chang2021explaining}.

\textbf{Macaque-Synthetic dataset.} This dataset \cite{majaj2015simple} is also about macaque neural responses which are recorded by electrodes under 3,200 synthetic image stimuli, and used for neural prediction in the initial version of Brain-Score \cite{schrimpf2020brain}. The image stimuli are generated by adding a 2D projection of a 3D object model to a natural background. The objects consist of eight categories, each with eight subclasses. The position, pose, and size of each object are randomly selected. 88 neurons of V4 and 168 neurons of IT are recorded. The neural responses are preprocessed to the form of average firing rate and can be downloaded from Brain-Score.

\subsection{Models}
\label{methods.model}

Since the core visual function of macaque and mouse visual cortex is to recognize objects, the basic premise of model selection is that the model has good performance on object recognition tasks (e.g. classification on ImageNet). Based on this premise, we employ 12 SNNs, 43 CNNs, and 26 vision transformers, all of which are pretrained on the ImageNet dataset and perform well in the classification task. As for SNNs, we use SEW ResNet as the base model, which is the deepest and SOTA directly trained SNN \cite{fang2021deep}. Furthermore, by combining the residual block used in SEW ResNet and the hierarchy of the visual cortex, we build several new SNNs and train them on the ImageNet using SpikingJelly \cite{SpikingJelly} (see Appendix A for model structures and the details of model training). As for CNNs and vision transformers, we use 44 models from the Torchvision model zoo \cite{paszke2019pytorch}, 22 models from the Timm model zoo \cite{rw2019timm} and 3 models from the brain-like CNNs, CORnet family \cite{kubilius2019brain}. In the feature extraction procedures of all models, we feed the same set of images used in biological experiments to the pretrained models and obtain features from all chosen layers. Different from CNNs and vision transformers, the features of SNNs are spikes in multiple time steps.

\subsection{Similarity Metrics}
\label{methods.metric}

To obtain the representation similarity between biological visual cortex and computational models, we apply three similarity metrics to computing similarity scores: representational similarity analysis (RSA) \cite{kriegeskorte2008matching, kriegeskorte2008representational}, regression-based encoding method \cite{carandini2005we, yamins2014performance, schrimpf2020brain, schrimpf2020integrative} and singular vector canonical correlation analysis (SVCCA) \cite{raghu2017svcca, morcos2018insights}. RSA has already been widely used to analyze neural representations of a model and a brain to different stimuli at the population level, while the regression-based encoding method directly fits the model features to neural activity data. SVCCA is originally proposed to compare features of deep neural networks, and then \cite{shi2019comparison} used it to compare representation matrices from mouse visual cortex and DNNs, which demonstrated its effectiveness.

With the same model and same cortical region, we use these metrics for a layer-by-layer comparison to compute the similarity scores. The maximum similarity score across layers for a given cortical region is considered to be the level of representation similarity between the model and the cortical region. Finally, in a given dataset, we take the average score of all cortical regions as the final similarity score for each model, which gives the overall model rankings. The implementation of each similarity metric is as follows. 

\textbf{RSA.} For two response matrices $R \in \mathbb{R}^{n \times m}$ from each layer of models and each cortical region, where $n$ is the number of units/neurons and $m$ is the number of stimuli, we calculate the representational similarity between the responses to each pair of image stimuli using the Pearson correlation coefficient $r$, yielding two representational dissimilarity matrices ($RDM \in \mathbb{R}^{m \times m}$, where each element is the correlation distance $1 - r$). Then, the Spearman rank correlation coefficient between the flattened upper triangles of these two matrices is the metric score.

\textbf{Regression-Based Encoding Method.} Firstly, we run truncated singular value decomposition (TSVD) to reduce the feature dimension of model layers to 40. Secondly, the features after dimensionality reduction are fitted to the representations of each neuron by ridge regression. Finally, we compute the Pearson correlation coefficient between the predicted and ground-truth representations of each neuron and take the mean of all correlation coefficients as the metric score. More specifically, we apply leave-one-out cross-validation to obtain predicted representations of each neuron. For simplicity, we name this method 'TSVD-Reg'.

\textbf{SVCCA.} For both the responses of model layers and cortical regions, we use TSVD to reduce the dimension of unit/neuron to 40, yielding two reduced representation matrices. Then we apply canonical correlation analysis (CCA) to these two matrices to obtain a vector of correlation coefficients (the length of the vector is 40). The metric score is the mean of the vector. Because of the invariance of CCA to affine transformations \cite{raghu2017svcca}, in this procedure, we only need to ensure that the stimulus dimension is consistent and aligned, even if the unit/neuron dimension is different. Dimensionality reduction plays an important role in this method to make the number of model features comparable to the number of neurons in cortical regions, since the former usually far exceeds the latter. In addition, dimensionality reduction helps to determine which features are important to the original data, while CCA suffers in important feature detection. Using just CCA performs badly, which has been proven by \cite{raghu2017svcca}.

\section{Results}
\label{results}

\subsection{Comparisons of Representation Similarity Scores between SNNs and Other Types of Models}
\label{results.overall}

\begin{figure}[t]
	\centering
	\includegraphics[width=0.48\textwidth]{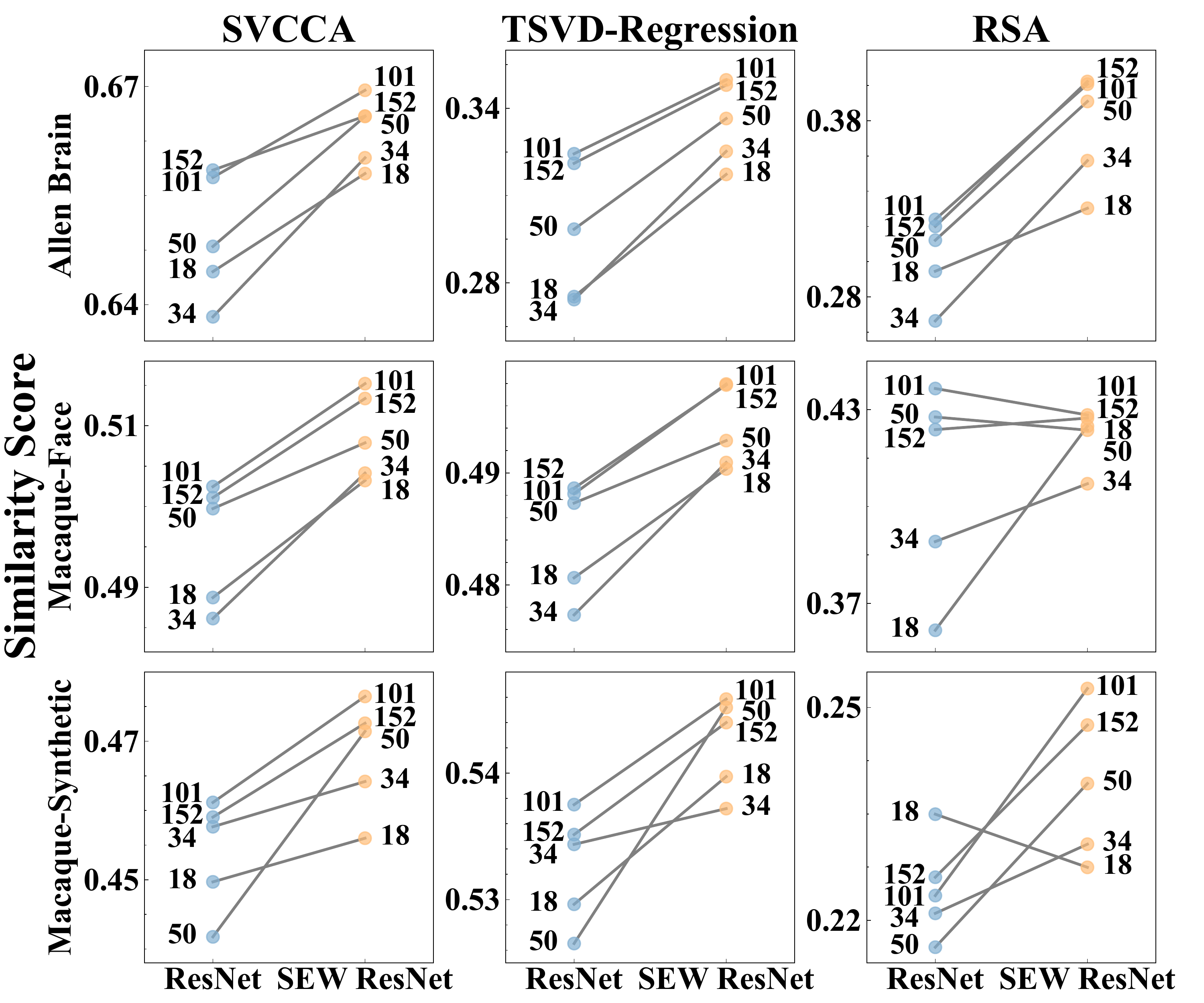}
	\caption{For three datasets and three similarity metrics, each point indicates the final representation similarity score of a model. Each pair of SEW ResNet and ResNet with the same depth are linked by a gray solid line. In almost all conditions, SEW ResNet outperforms ResNet by a large margin.}
	\label{Fig.resnet_compare}
\end{figure}

To check how similar the models are to the visual cortex’s mechanisms in visual processing, we rank the final similarity scores of all models and conduct comparisons among three types of models (CNNs, SNNs, and vision transformers). Specially, we focus on comparing SNN (SEW ResNet) and CNN (ResNet) with the same depth and almost the same architectures (Figure \ref{Fig.resnet_compare}). The final similarity score of a model is the average similarity score across all cortical regions. (The overall rankings can be found in Appendix B and the comparisons among three types of models are shown in Appendix C.)

\textbf{Allen brain mouse dataset.} No single model achieves the highest final similarity scores with all three metrics. For a fair comparison, we apply the paired t-test to SEW ResNet and ResNet with the same depth. For all three metrics, SEW ResNet performs better than ResNet by a large margin ($t=5.857$, $p=0.004$; $t=7.666$, $p=0.002$; $t=7.592$, $p=0.002$)\footnote{The results of the three similarity metrics are separated by semicolons, in the order of SVCCA, TSVD-Reg, and RSA. Other results that appear below also correspond to the three metrics in this order, unless the correspondence is stated in the text.}. 

\textbf{Macaque-Face dataset.} For both SVCCA and TSVD-Reg, Wide-SEW-ResNet14 and Wide-SEW-ResNet8 achieve the first and second highest final similarity scores respectively. But for RSA, TNT-S and Inception-ResNet-V2 take their place and outperform other models by a large margin. As for SEW ResNet and ResNet, the former performs significantly better than the latter for both SVCCA and TSVD-Reg ($t=8.195$, $p=0.001$; $t=7.528$, $p=0.002$). However, the difference is not significant for RSA ($t=1.117$, $p=0.327$). Specifically, the similarity score of SEW ResNet152 is only slightly higher than that of ResNet152, and at the depth of 50 and 101, SEW ResNet's scores are lower than ResNet's.

\textbf{Macaque-Synthetic dataset.} Similar to the results of Allen Brain dataset, no model performs best for all three metrics. SEW ResNet performs moderately better than ResNet ($t=3.354$, $p=0.028$; $t=3.824$, $p=0.019$; $t=2.343$, $p=0.079$). The only contrary is that SEW ResNet18 performs worse than ResNet18 for RSA.

\begin{figure}[t]
	\centering
	\includegraphics[width=0.48\textwidth]{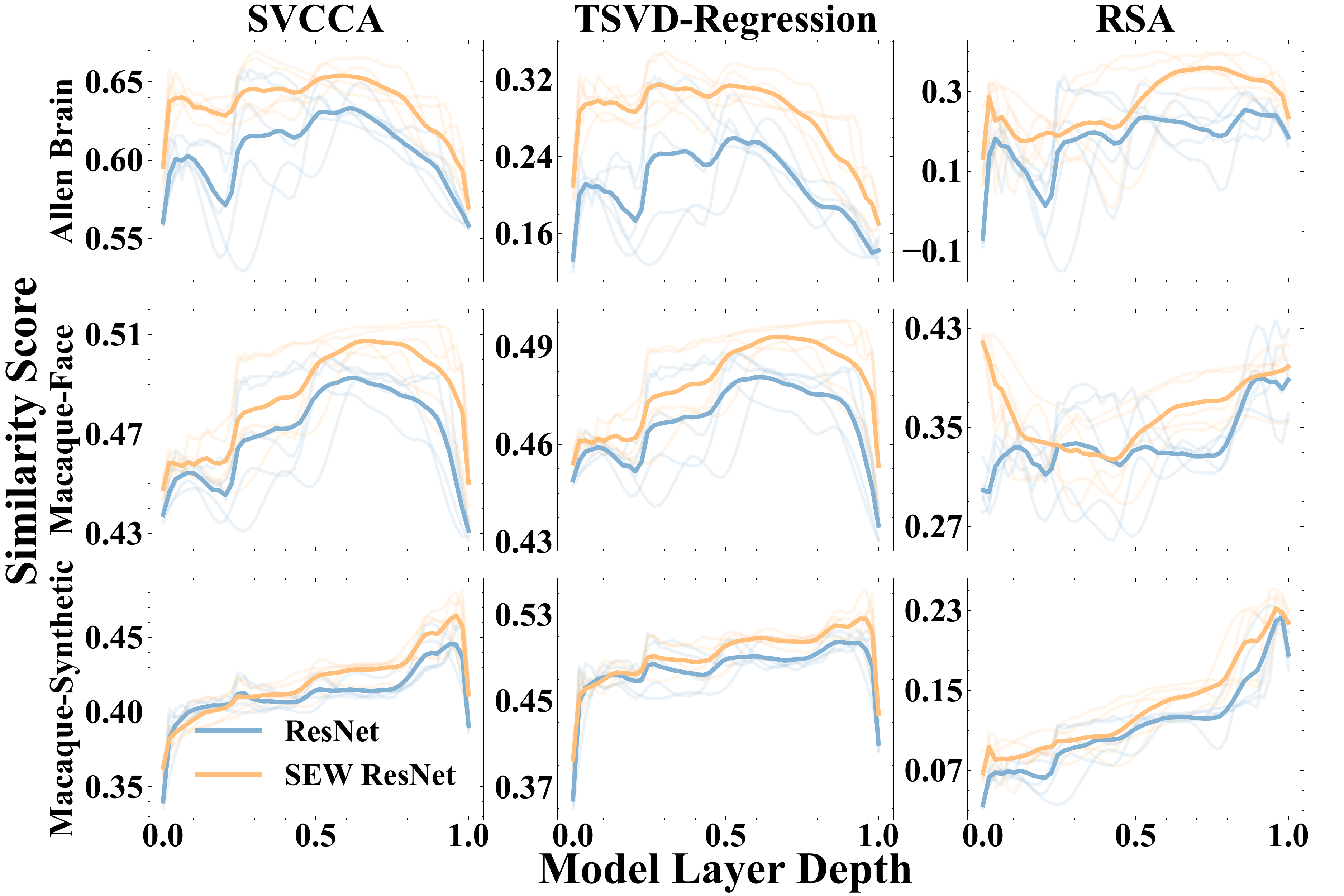}
	\caption{For three datasets and three similarity metrics, we plot the trajectories of similarity score with model layer depth. The models are divided into two groups: ResNet and SEW ResNet. The normalized layer depth ranges from 0 (the first layer) to 1 (the last layer). Because the depths of models are not the same, we first discretize the normalized depth into 50 bins, and then apply the cubic spline interpolation to the scores of each model, yielding the smooth trajectories shown in the plot. The fine, semitransparent lines are the trajectories of each model. The thick lines are the average trajectories among each group.}
	\label{Fig.resnet_trajectory_compare}
\end{figure}

Further, to check the details of comparison between the SNNs and their CNN counterparts, we analyze the trajectories of similarity score across model layers (Figure \ref{Fig.resnet_trajectory_compare}). As for ResNet and SEW ResNet with the same depth, the trends of their similarities across model layers are almost the same, but the former's trajectory is generally below the latter's. In other words, the similarity scores of SEW ResNet are higher than those of ResNet at almost all layers.

Taken together, the results suggest that when the overall architectures and depth are the same, SNNs with spiking neurons perform consistently better than their counterparts of CNNs with an average increase of $6.6\%$. Besides, SEW ResNet14 also outperforms the brain-like recurrent CNN, CORnet-S, with the same number of layers (see more details in Appendix B). Two properties of SNNs might contribute to the higher similarity scores. On the one hand, IF neurons are the basic neurons of spiking neural networks. The IF neuron uses several differential equations to roughly approximate the membrane potential dynamics of biological neurons, which provides a more biologically plausible spike mechanism for the network. On the other hand, the spiking neural network is able to capture the temporal features by incorporating both time and binary signals, just like the biological visual system during information processing.

\subsection{Best Layers across Cortical Regions Reveal Functional Hierarchy in the Visual Cortex of Macaques and Mice}

\begin{figure}[t]
	\centering
	\subfigure[Allen Brain mouse dataset]{\includegraphics[width=0.47\textwidth]{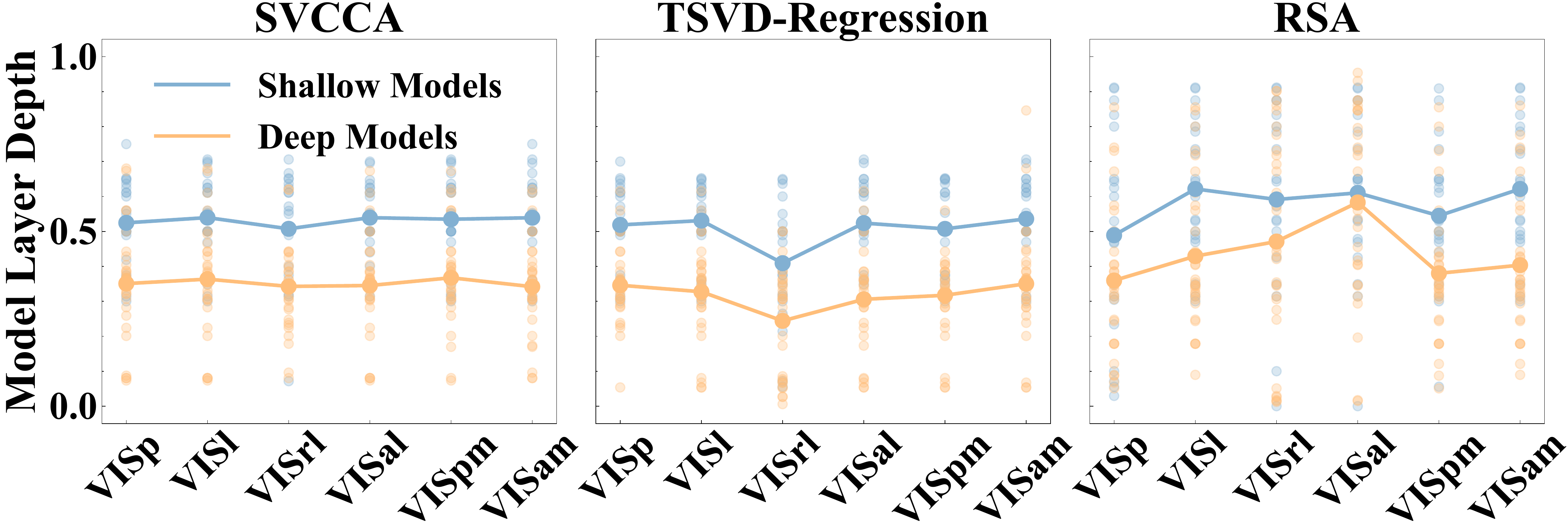}}
	\subfigure[Macaque-Face dataset and Macaque-Synthetic dataset]{\includegraphics[width=0.47\textwidth]{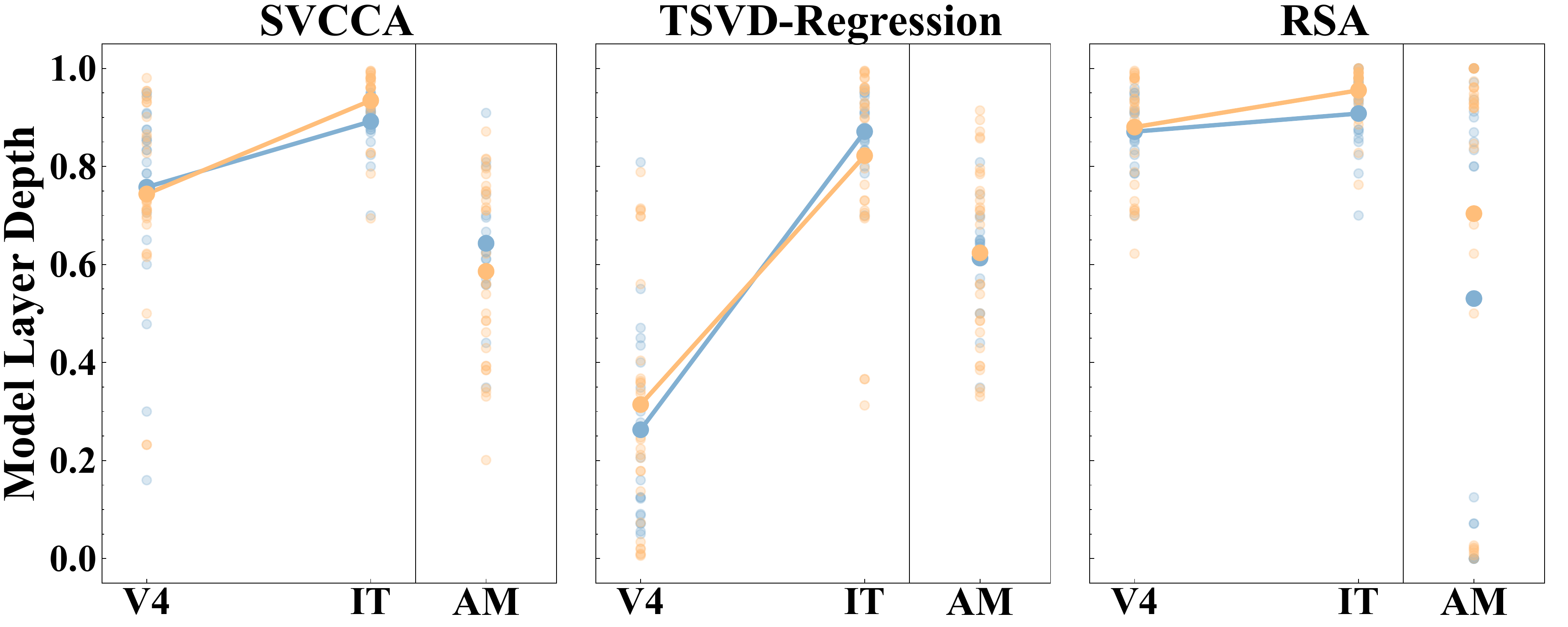}}
	\caption{For three datasets, we plot the normalized depth of the layer that achieves the top similarity score in each cortical region and each metric. Based on model depth, neural networks are divided into two groups: shallow models with less than 50 layers and deep models with more than 50 layers. The normalized layer depth ranges from 0 (the first layer) to 1 (the last layer). Each small point indicates an individual model. The large point indicates the average depth across a group.}
	\label{Fig.best_layer}
\end{figure}

To figure out the distinctions in the functional hierarchy between macaques and mice, for each cortical region, we obtain the normalized depth of the layer that achieves the highest similarity score in each model. Then, we divide models (excluding vision transformers) into two groups based on their depths and conduct investigations on these two groups separately. A nonparametric ANOVA is applied to each group for testing whether layer depths change significantly across cortical regions.

For mouse visual cortex (Figure \ref{Fig.best_layer} (a)), taking the deep model group as an example, ANOVA shows overall significant changes in depth across cortical regions for TSVD-Reg and RSA (Friedman's ${\chi}^2=49.169$, $p=2.0 \times 10^{-9}$; ${\chi}^2=19.455$, $p=0.002$). But there is no significant change for SVCCA (${\chi}^2=8.689$, $p=0.122$). According to these results, the differences in depth across regions are indeterminacy and irregular. Meanwhile, the trends of layer depth between some regions contradict the hierarchy observed in physiological experiments of mice (those between VISp and VISrl for TSVD-Reg and between VISal and VISpm for RSA). However, for macaque visual cortex (Figure \ref{Fig.best_layer} (b)), there are significant differences ($t=-5.451$, $p=6.5 \times 10^{-6}$; $t=-8.312$, $p=2.8 \times 10^{-9}$; $t=-3.782$, $p=6.9 \times 10^{-4}$, also taking the deep model group as an example) between V4 and IT, and the trend is consistent with the information processing hierarchy in primate visual cortex.

The comparative analyses of the best layer depths of the shallow and deep model groups also exhibit the differences between macaques and mice. For mouse visual cortex, the best layer depths of shallow models are significantly higher than those of deep models. Compared to deep models, most shallow models achieve the top similarity scores in intermediate and even later layers. Differently, for macaque visual cortex, the depth of models has little effect on the depth of the most similar layer. What's more, we find that the most similar layer of mouse visual cortex always occurs after the $28 \times 28$ feature map is downsampled to $14 \times 14$, which leads to the layer depths' difference between shallow and deep models. Nevertheless, the best layer of macaque IT appears in the last part of networks, where the feature map has been downsampled more times.

\begin{table*}[t]
	\centering
	\begin{tabular}{llll}
		\toprule
		\diagbox{\textbf{Dataset}}{\textbf{Metric}} & \textbf{SVCCA} & \textbf{TSVD-Regression} & \textbf{RSA} \\
		\midrule
		\textbf{Allen Brain mouse dataset} & \makecell[l]{$r=-0.654$,\\$p=2.0 \times 10^{-6}$} & \makecell[l]{$r=-0.596$,\\$p=2.4 \times 10^{-5}$} & \makecell[l]{$r=-0.548$,\\$p=1.4 \times 10^{-4}$} \\
		\midrule
		\textbf{Macaque-Face dataset} & --- & --- & --- \\
		\midrule
		\textbf{Macaque-Synthetic dataset} & --- & --- & --- \\
		\bottomrule
	\end{tabular}
	\caption{The correlation between the similarity scores and the number of parameters. $r$ is Spearman's rank correlation coefficient. "---" indicates that there is no significant correlation.}
	\label{table.parameters}
\end{table*}

\begin{table*}[t]
	\centering
	\begin{tabular}{llll}
		\toprule
		\diagbox{\textbf{Dataset}}{\textbf{Metric}} & \textbf{SVCCA} & \textbf{TSVD-Regression} & \textbf{RSA} \\
		\midrule
		\textbf{Allen Brain mouse dataset} & --- & --- & --- \\
		\midrule
		\textbf{Macaque-Face dataset} & \makecell[l]{$r=0.657$,\\$p=4.2 \times 10^{-6}$} & \makecell[l]{$r=0.634$,\\$p=1.1 \times 10^{-5}$} & \makecell[l]{$r=0.527$,\\$p=4.7 \times 10^{-4}$} \\
		\midrule
		\textbf{Macaque-Synthetic dataset} & --- & \makecell[l]{$r=-0.408$,\\$p=0.009$} & \makecell[l]{$r=-0.575$,\\$p=1.1 \times 10^{-4}$} \\
		\bottomrule
	\end{tabular}
	\caption{The correlation between the similarity scores and the model depth. $r$ is Spearman's rank correlation coefficient. "---" indicates that there is no significant correlation.}
	\label{table.depth}
\end{table*}

In summary, our results might reveal two distinctions in the functional hierarchy between macaques and mice. First, there is a distinct functional hierarchical structure of macaque ventral visual pathway, while there might be no clear sequential functional hierarchy in mouse visual cortex. One explanation is that the mouse visual cortex is organized into a parallel structure and the function of mouse cortical regions are more generalized and homogeneous than those of macaques. Another possibility would be that even though the sequential relations exist among mouse cortical regions as proposed in anatomical and physiological work, they are too weak for the current deep neural networks to capture. Additionally, mice perform more complex visual tasks than expected with a limited brain capacity \cite{djurdjevic2018accuracy}. Consequently, the neural responses of mouse visual cortex may contain more information not related to object recognition that neural networks focus on. Secondly, it is well known that the units in the neural networks get larger receptive fields after downsampling, and through the analyses of differences between two groups of models based on depth, we find the feature map of the best layer for mouse is downsampled fewer times than that for macaque. Based on these results, we provide computational evidence that the increased ratio of the receptive field size in cortical regions across the mouse visual pathway is smaller than those across the macaque visual pathways, which echoes some physiological work \cite{siegle2021survey, zhu2013multi}.

\subsection{Structures and Mechanisms of Models Reveal Processing Mechanisms in the Visual Cortex of Macaques and Mice}

To explore the processing mechanisms in the visual cortex of macaques and mice, we investigate the model properties from the whole to the details. As shown in Table \ref{table.parameters} and \ref{table.depth}, we first measure the correlation between the similarity scores and the sizes (i.e. the number of trainable parameters and the depth) of network models. For Allen Brain mouse dataset, there are significant negative correlations between the similarity scores and the number of parameters for three metrics while there is no correlation with the depth. Conversely, for the two macaque neural datasets, the similarity scores are highly correlated with the depth of networks, but not with the number of parameters. Specifically, there is a positive correlation for Macaque-Face dataset while a negative correlation for Macaque-Synthetic dataset. (We also apply the linear regression to analyze the correlation between the similarity scores and the model size. The results are consistent with Spearman's rank correlation and are shown in Appendix E). Based on these results, we further investigate more detailed properties of neural networks to explain the processing mechanisms in the visual cortex.

For the mouse dataset, on the one hand, the best layer depths show non-significant changes across the mouse cortical regions as mentioned in the previous section. On the other hand, the similarity scores of the mouse dataset are only correlated with the number of model parameters but not with the depth of models. It calls into the question whether any detailed structures in the neural networks help to reduce the number of parameters and improve its similarity to mouse visual cortex. Therefore, we explore the commonalities between models that have the top 20\% representation similarities (see Appendix D) for Allen Brain dataset. As expected, the top models contain similar structures, such as fire module, inception module, and depthwise separable convolution. All these structures essentially process information through multiple branches/channels and then integrate the features from each branch. The models with this type of structure outperform other models ($t=2.411$, $p=0.024$; $t=3.030$, $p=0.007$; $t=1.174$, $p=0.247$). Moreover, we apply the depthwise separable convolution to SNNs, which yields a positive effect. The representation similarity of Spiking-MobileNet is higher than SEW-ResNet50 with a similar depth (+0.8\%; +3.9\%; +12.1\%). In fact, some studies using multiple pathways simulate the functions of mouse visual cortex to some extent \cite{shi2022mousenet, nayebi2022mouse}. Our results further suggest that not only the mouse visual cortex might be an organization of parallel structures, but also there are extensive parallel information processing streams between each pair of cortical regions \cite{wang2012network, siegle2021survey}.

\begin{figure*}[t]
	\centering
	\includegraphics[width=0.99\textwidth]{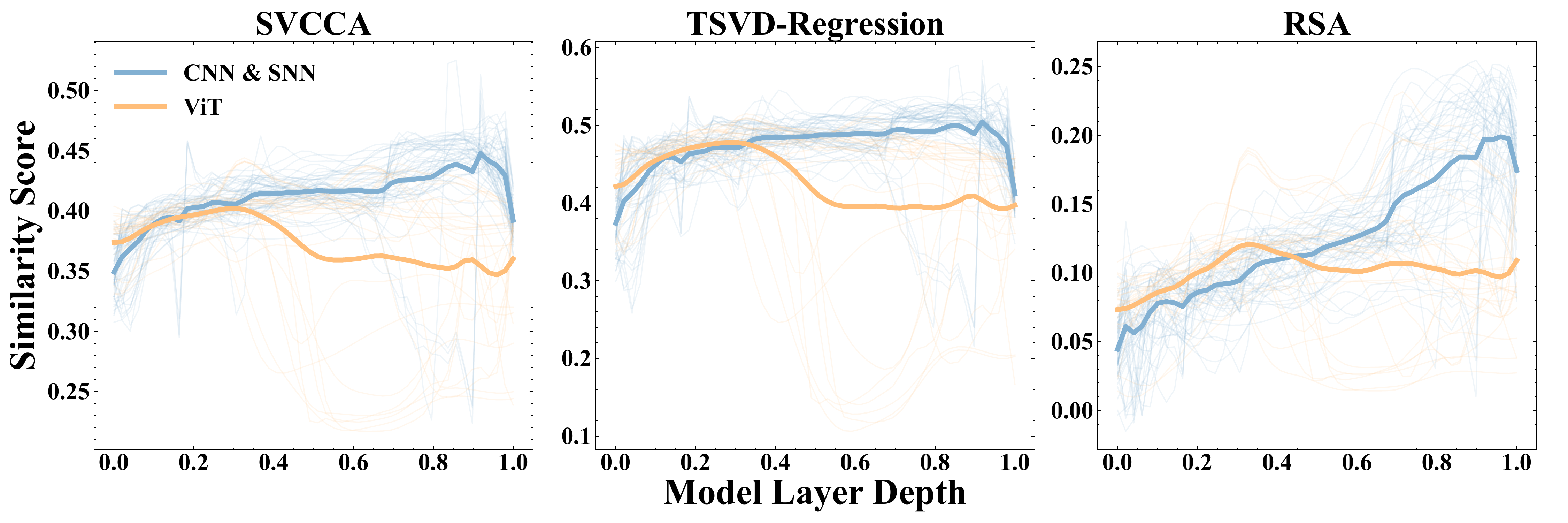}
	\caption{For Macaque-Synthetic dataset, trajectories of similarity score with model layer depth are plotted. The models are divided into two groups: ViT and CNN\&SNN. The normalized layer depth ranges from 0 (the first layer) to 1 (the last layer). The calculation and plotting of the trajectories are the same as Figure \ref{Fig.resnet_trajectory_compare}.}
	\label{Fig.trajectory_compare}
\end{figure*}

For the two macaque datasets with different stimuli, not only are the model rankings significantly different, but also the correlations between the similarity scores and the model depth are totally opposite. These results corroborate the following two processing mechanisms in macaques: the ventral visual stream of primate visual cortex possesses canonical coding principles at different stages; the brain exhibits a high degree of functional specialization, such as the visual recognition of faces and other objects, which is reflected in the different neural responses of the corresponding region (although the face patch AM is a sub-network of IT, they differ in the neural representations). Besides, as shown in Figure \ref{Fig.trajectory_compare}, the similarity scores of vision transformers reach the maximum in the early layers and then decrease. Differently, the scores of CNNs and SNNs keep trending upwards, reaching the maximum in almost the last layer. On the other hand, Appendix C shows that vision transformers perform well in Macaque-Face dataset but poorly in Macaque-Synthetic dataset. Considering the features extraction mechanism of vision transformers, it divides the image into several patches and encodes each patch as well as their internal relation by self-attention. This mechanism is effective for face images that are full of useful information. However, the synthetic image consists of a central target object and a naturalistic background. When vision transformers are fed with this type of stimuli, premature integration of global information can lead to model representations containing noise from the unrelated background. What's more, when we take all models with the top 20\% representation similarities as a whole for analyses, as described in the above paragraph, the properties that enable networks to achieve higher neural similarity are not yet clear. Taken together, the computational mechanism of the better models may reveal core processing divergence to different types of stimuli in the visual cortex.

\section{Discussion}

In this work, we take large-scale neural representation similarity experiments as a basis, aided by analyses of the similarities across models and the visual cortical regions. Compared to other work, we introduce SNNs in the similarity analyses with biological neural responses for the first time, showing that SNNs achieve higher similarity scores than CNNs that have the same depth and almost the same architectures. As analyzed in Section 3.1, two properties of SNNs might serve as the explanations for their high similarity scores. 

The subsequent analyses of the models' simulation performance and structures indicate significant differences in functional hierarchies between macaque and mouse visual cortex. As for macaques, we observed a clear sequential hierarchy. However, as for mouse visual cortex, some work \cite{conwell2021neural} exhibits that the trend of the model feature complexity roughly matches the processing hierarchy, but other work suggests that the cortex \cite{shi2019comparison, nayebi2022mouse} is organized into a parallel structure. Our results are more supportive of the latter. Furthermore, we provide computational evidence not only that the increased ratio of the receptive field size in cortical regions across the mouse visual pathway is smaller than those across the macaque visual pathway, but also that there may be multiple pathways with parallel processing streams between mouse cortical regions. Our results also clearly reveal that the processing mechanisms of macaque visual cortex differ to various stimuli. These findings provide us with new insights into the visual processing mechanisms of macaque and mouse, which are the two species that dominate the research of biological vision systems and differ considerably from each other.

Compared to CNNs, the study of task-driven deep SNNs is just in its initial state. Although we demonstrate that SNNs outperform their counterparts of CNNs, SNNs exhibit similar properties as CNNs in the further analyses. In this work, we only build several new SNNs by taking the hints from the biological visual hierarchy, while many well-established structures and learning algorithms in CNNs have not been applied to SNNs yet. In addition, the neural datasets used in our experiments are all collected under static image stimuli, lacking rich dynamic information to some certain, which may not fully exploit the properties of SNNs. Given that SNNs perform well in the current experiments, we hope to explore more potential of SNNs in future work.

In conclusion, as more biologically plausible neural networks, SNNs may serve as a shortcut to explore the biological visual cortex. With studies on various aspects of SNNs, such as model architectures, learning algorithms, processing mechanisms, and neural coding methods, it's highly promising to better explain the sophisticated, complex, and diverse vision systems in the future.

\section{Ethics Statement}

The biological neural datasets used in our experiments are obtained from public datasets or from published papers with the authors’ consent.

\section{Acknowledgements}

We thank L. Chang for providing Macaque-Face dataset. This work is supported by the National Natural Science Foundation of China (No.61825101, No.62027804, and No.62088102).

\bibliography{ref}

\begin{thebibliography}{54}
\providecommand{\natexlab}[1]{#1}

\bibitem[{Bakhtiari et~al.(2021)Bakhtiari, Mineault, Lillicrap, Pack, and
  Richards}]{bakhtiari2021functional}
Bakhtiari, S.; Mineault, P.; Lillicrap, T.; Pack, C.; and Richards, B. 2021.
\newblock The functional specialization of visual cortex emerges from training
  parallel pathways with self-supervised predictive learning.
\newblock In \emph{Advances in Neural Information Processing Systems 34},
  25164--25178.

\bibitem[{Bashivan, Kar, and DiCarlo(2019)}]{bashivan2019neural}
Bashivan, P.; Kar, K.; and DiCarlo, J.~J. 2019.
\newblock Neural population control via deep image synthesis.
\newblock \emph{Science}, 364(6439): eaav9436.

\bibitem[{Brette et~al.(2007)Brette, Rudolph, Carnevale, Hines, Beeman, Bower,
  Diesmann, Morrison, Goodman, Harris et~al.}]{brette2007simulation}
Brette, R.; Rudolph, M.; Carnevale, T.; Hines, M.; Beeman, D.; Bower, J.~M.;
  Diesmann, M.; Morrison, A.; Goodman, P.~H.; Harris, F.~C.; et~al. 2007.
\newblock Simulation of networks of spiking neurons: a review of tools and
  strategies.
\newblock \emph{Journal of computational neuroscience}, 23(3): 349--398.

\bibitem[{Cadena et~al.(2019)Cadena, Sinz, Muhammad, Froudarakis, Cobos,
  Walker, Reimer, Bethge, Tolias, and Ecker}]{cadena2019well}
Cadena, S.~A.; Sinz, F.~H.; Muhammad, T.; Froudarakis, E.; Cobos, E.; Walker,
  E.~Y.; Reimer, J.; Bethge, M.; Tolias, A.; and Ecker, A.~S. 2019.
\newblock How well do deep neural networks trained on object recognition
  characterize the mouse visual system?
\newblock In \emph{NeurIPS Neuro AI Workshop}.

\bibitem[{Carandini et~al.(2005)Carandini, Demb, Mante, Tolhurst, Dan,
  Olshausen, Gallant, and Rust}]{carandini2005we}
Carandini, M.; Demb, J.~B.; Mante, V.; Tolhurst, D.~J.; Dan, Y.; Olshausen,
  B.~A.; Gallant, J.~L.; and Rust, N.~C. 2005.
\newblock Do we know what the early visual system does?
\newblock \emph{Journal of Neuroscience}, 25(46): 10577--10597.

\bibitem[{Chang et~al.(2021)Chang, Egger, Vetter, and
  Tsao}]{chang2021explaining}
Chang, L.; Egger, B.; Vetter, T.; and Tsao, D.~Y. 2021.
\newblock Explaining face representation in the primate brain using different
  computational models.
\newblock \emph{Current Biology}, 31(13): 2785--2795.

\bibitem[{Conwell et~al.(2021)Conwell, Mayo, Barbu, Buice, Alvarez, and
  Katz}]{conwell2021neural}
Conwell, C.; Mayo, D.; Barbu, A.; Buice, M.; Alvarez, G.; and Katz, B. 2021.
\newblock Neural regression, representational similarity, model zoology \&
  neural taskonomy at scale in rodent visual cortex.
\newblock In \emph{Advances in Neural Information Processing Systems 34},
  5590--5607.

\bibitem[{Dapello et~al.(2020)Dapello, Marques, Schrimpf, Geiger, Cox, and
  DiCarlo}]{dapello2020simulating}
Dapello, J.; Marques, T.; Schrimpf, M.; Geiger, F.; Cox, D.; and DiCarlo, J.~J.
  2020.
\newblock Simulating a primary visual cortex at the front of CNNs improves
  robustness to image perturbations.
\newblock In \emph{Advances in Neural Information Processing Systems 33},
  13073--13087.

\bibitem[{de~Vries et~al.(2020)de~Vries, Lecoq, Buice, Groblewski, Ocker,
  Oliver, Feng, Cain, Ledochowitsch, Millman et~al.}]{de2020large}
de~Vries, S.~E.; Lecoq, J.~A.; Buice, M.~A.; Groblewski, P.~A.; Ocker, G.~K.;
  Oliver, M.; Feng, D.; Cain, N.; Ledochowitsch, P.; Millman, D.; et~al. 2020.
\newblock A large-scale standardized physiological survey reveals functional
  organization of the mouse visual cortex.
\newblock \emph{Nature neuroscience}, 23(1): 138--151.

\bibitem[{Djurdjevic et~al.(2018)Djurdjevic, Ansuini, Bertolini, Macke, and
  Zoccolan}]{djurdjevic2018accuracy}
Djurdjevic, V.; Ansuini, A.; Bertolini, D.; Macke, J.~H.; and Zoccolan, D.
  2018.
\newblock Accuracy of rats in discriminating visual objects is explained by the
  complexity of their perceptual strategy.
\newblock \emph{Current biology}, 28(7): 1005--1015.

\bibitem[{Fang et~al.(2020)Fang, Chen, Ding, Chen, Yu, Zhou, Masquelier, Tian,
  and other contributors}]{SpikingJelly}
Fang, W.; Chen, Y.; Ding, J.; Chen, D.; Yu, Z.; Zhou, H.; Masquelier, T.; Tian,
  Y.; and other contributors. 2020.
\newblock SpikingJelly.
\newblock \url{https://github.com/fangwei123456/spikingjelly}.
\newblock Accessed: 2022-07-06.

\bibitem[{Fang et~al.(2021{\natexlab{a}})Fang, Yu, Chen, Huang, Masquelier, and
  Tian}]{fang2021deep}
Fang, W.; Yu, Z.; Chen, Y.; Huang, T.; Masquelier, T.; and Tian, Y.
  2021{\natexlab{a}}.
\newblock Deep residual learning in spiking neural networks.
\newblock In \emph{Advances in Neural Information Processing Systems 34},
  21056--21069.

\bibitem[{Fang et~al.(2021{\natexlab{b}})Fang, Yu, Chen, Masquelier, Huang, and
  Tian}]{fang2021incorporating}
Fang, W.; Yu, Z.; Chen, Y.; Masquelier, T.; Huang, T.; and Tian, Y.
  2021{\natexlab{b}}.
\newblock Incorporating learnable membrane time constant to enhance learning of
  spiking neural networks.
\newblock In \emph{Proceedings of the IEEE/CVF International Conference on
  Computer Vision (ICCV)}, 2661--2671.

\bibitem[{Gerstner and Kistler(2002)}]{gerstner2002spiking}
Gerstner, W.; and Kistler, W.~M. 2002.
\newblock \emph{Spiking neuron models: Single neurons, populations,
  plasticity}.
\newblock Cambridge university press.

\bibitem[{G{\"u}{\c{c}}l{\"u} and van Gerven(2015)}]{gucclu2015deep}
G{\"u}{\c{c}}l{\"u}, U.; and van Gerven, M.~A. 2015.
\newblock Deep neural networks reveal a gradient in the complexity of neural
  representations across the ventral stream.
\newblock \emph{Journal of Neuroscience}, 35(27): 10005--10014.

\bibitem[{He et~al.(2016)He, Zhang, Ren, and Sun}]{he2016deep}
He, K.; Zhang, X.; Ren, S.; and Sun, J. 2016.
\newblock Deep residual learning for image recognition.
\newblock In \emph{Proceedings of the IEEE Conference on Computer Vision and
  Pattern Recognition (CVPR)}, 770--778.

\bibitem[{Higgins et~al.(2021)Higgins, Chang, Langston, Hassabis, Summerfield,
  Tsao, and Botvinick}]{higgins2021unsupervised}
Higgins, I.; Chang, L.; Langston, V.; Hassabis, D.; Summerfield, C.; Tsao, D.;
  and Botvinick, M. 2021.
\newblock Unsupervised deep learning identifies semantic disentanglement in
  single inferotemporal face patch neurons.
\newblock \emph{Nature communications}, 12(1): 1--14.

\bibitem[{Hodgkin and Huxley(1952)}]{hodgkin1952quantitative}
Hodgkin, A.~L.; and Huxley, A.~F. 1952.
\newblock A quantitative description of membrane current and its application to
  conduction and excitation in nerve.
\newblock \emph{The Journal of physiology}, 117(4): 500.

\bibitem[{Howard et~al.(2017)Howard, Zhu, Chen, Kalenichenko, Wang, Weyand,
  Andreetto, and Adam}]{howard2017mobilenets}
Howard, A.~G.; Zhu, M.; Chen, B.; Kalenichenko, D.; Wang, W.; Weyand, T.;
  Andreetto, M.; and Adam, H. 2017.
\newblock Mobilenets: Efficient convolutional neural networks for mobile vision
  applications.
\newblock arXiv:1704.04861.

\bibitem[{Hubel and Wiesel(1959)}]{hubel1959receptive}
Hubel, D.~H.; and Wiesel, T.~N. 1959.
\newblock Receptive fields of single neurones in the cat's striate cortex.
\newblock \emph{The Journal of physiology}, 148(3): 574.

\bibitem[{Hubel and Wiesel(1962)}]{hubel1962receptive}
Hubel, D.~H.; and Wiesel, T.~N. 1962.
\newblock Receptive fields, binocular interaction and functional architecture
  in the cat's visual cortex.
\newblock \emph{The Journal of physiology}, 160(1): 106.

\bibitem[{Izhikevich(2004)}]{izhikevich2004model}
Izhikevich, E.~M. 2004.
\newblock Which model to use for cortical spiking neurons?
\newblock \emph{IEEE Transactions on Neural Networks}, 15(5): 1063--1070.

\bibitem[{Kasabov et~al.(2013)Kasabov, Dhoble, Nuntalid, and
  Indiveri}]{kasabov2013dynamic}
Kasabov, N.; Dhoble, K.; Nuntalid, N.; and Indiveri, G. 2013.
\newblock Dynamic evolving spiking neural networks for on-line spatio-and
  spectro-temporal pattern recognition.
\newblock \emph{Neural Networks}, 41: 188--201.

\bibitem[{Khaligh-Razavi and Kriegeskorte(2014)}]{khaligh2014deep}
Khaligh-Razavi, S.-M.; and Kriegeskorte, N. 2014.
\newblock Deep supervised, but not unsupervised, models may explain IT cortical
  representation.
\newblock \emph{PLoS computational biology}, 10(11): e1003915.

\bibitem[{Kietzmann et~al.(2019)Kietzmann, Spoerer, S{\"o}rensen, Cichy, Hauk,
  and Kriegeskorte}]{kietzmann2019recurrence}
Kietzmann, T.~C.; Spoerer, C.~J.; S{\"o}rensen, L.~K.; Cichy, R.~M.; Hauk, O.;
  and Kriegeskorte, N. 2019.
\newblock Recurrence is required to capture the representational dynamics of
  the human visual system.
\newblock \emph{Proceedings of the National Academy of Sciences}, 116(43):
  21854--21863.

\bibitem[{Kriegeskorte, Mur, and
  Bandettini(2008)}]{kriegeskorte2008representational}
Kriegeskorte, N.; Mur, M.; and Bandettini, P.~A. 2008.
\newblock Representational similarity analysis-connecting the branches of
  systems neuroscience.
\newblock \emph{Frontiers in systems neuroscience}, 2: 4.

\bibitem[{Kriegeskorte et~al.(2008)Kriegeskorte, Mur, Ruff, Kiani, Bodurka,
  Esteky, Tanaka, and Bandettini}]{kriegeskorte2008matching}
Kriegeskorte, N.; Mur, M.; Ruff, D.~A.; Kiani, R.; Bodurka, J.; Esteky, H.;
  Tanaka, K.; and Bandettini, P.~A. 2008.
\newblock Matching categorical object representations in inferior temporal
  cortex of man and monkey.
\newblock \emph{Neuron}, 60(6): 1126--1141.

\bibitem[{Kubilius et~al.(2019)Kubilius, Schrimpf, Kar, Rajalingham, Hong,
  Majaj, Issa, Bashivan, Prescott-Roy, Schmidt et~al.}]{kubilius2019brain}
Kubilius, J.; Schrimpf, M.; Kar, K.; Rajalingham, R.; Hong, H.; Majaj, N.;
  Issa, E.; Bashivan, P.; Prescott-Roy, J.; Schmidt, K.; et~al. 2019.
\newblock Brain-like object recognition with high-performing shallow recurrent
  ANNs.
\newblock In \emph{Advances in Neural Information Processing Systems 32},
  12785--12796.

\bibitem[{LeCun, Bengio, and Hinton(2015)}]{lecun2015deep}
LeCun, Y.; Bengio, Y.; and Hinton, G. 2015.
\newblock Deep learning.
\newblock \emph{nature}, 521(7553): 436--444.

\bibitem[{Maass(1997)}]{maass1997networks}
Maass, W. 1997.
\newblock Networks of spiking neurons: the third generation of neural network
  models.
\newblock \emph{Neural networks}, 10(9): 1659--1671.

\bibitem[{Majaj et~al.(2015)Majaj, Hong, Solomon, and
  DiCarlo}]{majaj2015simple}
Majaj, N.~J.; Hong, H.; Solomon, E.~A.; and DiCarlo, J.~J. 2015.
\newblock Simple learned weighted sums of inferior temporal neuronal firing
  rates accurately predict human core object recognition performance.
\newblock \emph{Journal of Neuroscience}, 35(39): 13402--13418.

\bibitem[{Morcos, Raghu, and Bengio(2018)}]{morcos2018insights}
Morcos, A.; Raghu, M.; and Bengio, S. 2018.
\newblock Insights on representational similarity in neural networks with
  canonical correlation.
\newblock In \emph{Advances in Neural Information Processing Systems 31},
  5732--5741.

\bibitem[{Nayebi et~al.(2018)Nayebi, Bear, Kubilius, Kar, Ganguli, Sussillo,
  DiCarlo, and Yamins}]{nayebi2018task}
Nayebi, A.; Bear, D.; Kubilius, J.; Kar, K.; Ganguli, S.; Sussillo, D.;
  DiCarlo, J.~J.; and Yamins, D.~L. 2018.
\newblock Task-driven convolutional recurrent models of the visual system.
\newblock In \emph{Advances in Neural Information Processing Systems 31},
  5295--5306.

\bibitem[{Nayebi et~al.(2022)Nayebi, Kong, Zhuang, Gardner, Norcia, and
  Yamins}]{nayebi2022mouse}
Nayebi, A.; Kong, N.~C.; Zhuang, C.; Gardner, J.~L.; Norcia, A.~M.; and Yamins,
  D. L.~K. 2022.
\newblock Mouse visual cortex as a limited resource system that self-learns an
  ecologically-general representation.
\newblock \emph{bioRxiv}.

\bibitem[{Neftci, Mostafa, and Zenke(2019)}]{neftci2019surrogate}
Neftci, E.~O.; Mostafa, H.; and Zenke, F. 2019.
\newblock Surrogate gradient learning in spiking neural networks: Bringing the
  power of gradient-based optimization to spiking neural networks.
\newblock \emph{IEEE Signal Processing Magazine}, 36(6): 51--63.

\bibitem[{Paszke et~al.(2019)Paszke, Gross, Massa, Lerer, Bradbury, Chanan,
  Killeen, Lin, Gimelshein, Antiga et~al.}]{paszke2019pytorch}
Paszke, A.; Gross, S.; Massa, F.; Lerer, A.; Bradbury, J.; Chanan, G.; Killeen,
  T.; Lin, Z.; Gimelshein, N.; Antiga, L.; et~al. 2019.
\newblock Pytorch: An imperative style, high-performance deep learning library.
\newblock In \emph{Advances in Neural Information Processing Systems 32},
  8024–--8035.

\bibitem[{Pham, Packianather, and Charles(2008)}]{pham2008control}
Pham, D.~T.; Packianather, M.~S.; and Charles, E. 2008.
\newblock Control chart pattern clustering using a new self-organizing spiking
  neural network.
\newblock \emph{Proceedings of the Institution of Mechanical Engineers, Part B:
  Journal of Engineering Manufacture}, 222(10): 1201--1211.

\bibitem[{Ponce et~al.(2019)Ponce, Xiao, Schade, Hartmann, Kreiman, and
  Livingstone}]{ponce2019evolving}
Ponce, C.~R.; Xiao, W.; Schade, P.~F.; Hartmann, T.~S.; Kreiman, G.; and
  Livingstone, M.~S. 2019.
\newblock Evolving images for visual neurons using a deep generative network
  reveals coding principles and neuronal preferences.
\newblock \emph{Cell}, 177(4): 999--1009.

\bibitem[{Raghu et~al.(2017)Raghu, Gilmer, Yosinski, and
  Sohl-Dickstein}]{raghu2017svcca}
Raghu, M.; Gilmer, J.; Yosinski, J.; and Sohl-Dickstein, J. 2017.
\newblock {SVCCA:} Singular vector canonical correlation analysis for deep
  learning dynamics and interpretability.
\newblock In \emph{Advances in Neural Information Processing Systems 30},
  6076--6085.

\bibitem[{Sandler et~al.(2018)Sandler, Howard, Zhu, Zhmoginov, and
  Chen}]{sandler2018mobilenetv2}
Sandler, M.; Howard, A.; Zhu, M.; Zhmoginov, A.; and Chen, L.-C. 2018.
\newblock Mobilenetv2: Inverted residuals and linear bottlenecks.
\newblock In \emph{Proceedings of the IEEE Conference on Computer Vision and
  Pattern Recognition (CVPR)}, 4510--4520.

\bibitem[{Schrimpf et~al.(2020{\natexlab{a}})Schrimpf, Kubilius, Hong, Majaj,
  Rajalingham, Issa, Kar, Bashivan, Prescott-Roy, Geiger
  et~al.}]{schrimpf2020brain}
Schrimpf, M.; Kubilius, J.; Hong, H.; Majaj, N.~J.; Rajalingham, R.; Issa,
  E.~B.; Kar, K.; Bashivan, P.; Prescott-Roy, J.; Geiger, F.; et~al.
  2020{\natexlab{a}}.
\newblock Brain-score: Which artificial neural network for object recognition
  is most brain-like?
\newblock \emph{bioRxiv}.

\bibitem[{Schrimpf et~al.(2020{\natexlab{b}})Schrimpf, Kubilius, Lee, Murty,
  Ajemian, and DiCarlo}]{schrimpf2020integrative}
Schrimpf, M.; Kubilius, J.; Lee, M.~J.; Murty, N. A.~R.; Ajemian, R.; and
  DiCarlo, J.~J. 2020{\natexlab{b}}.
\newblock Integrative benchmarking to advance neurally mechanistic models of
  human intelligence.
\newblock \emph{Neuron}, 108(3): 413--423.

\bibitem[{Shi, Shea-Brown, and Buice(2019)}]{shi2019comparison}
Shi, J.; Shea-Brown, E.; and Buice, M. 2019.
\newblock Comparison against task driven artificial neural networks reveals
  functional properties in mouse visual cortex.
\newblock In \emph{Advances in Neural Information Processing Systems 32},
  5765--5775.

\bibitem[{Shi et~al.(2022)Shi, Tripp, Shea-Brown, Mihalas, and
  A.~Buice}]{shi2022mousenet}
Shi, J.; Tripp, B.; Shea-Brown, E.; Mihalas, S.; and A.~Buice, M. 2022.
\newblock MouseNet: A biologically constrained convolutional neural network
  model for the mouse visual cortex.
\newblock \emph{PLOS Computational Biology}, 18(9): e1010427.

\bibitem[{Siegle et~al.(2021)Siegle, Jia, Durand, Gale, Bennett, Graddis,
  Heller, Ramirez, Choi, Luviano et~al.}]{siegle2021survey}
Siegle, J.~H.; Jia, X.; Durand, S.; Gale, S.; Bennett, C.; Graddis, N.; Heller,
  G.; Ramirez, T.~K.; Choi, H.; Luviano, J.~A.; et~al. 2021.
\newblock Survey of spiking in the mouse visual system reveals functional
  hierarchy.
\newblock \emph{Nature}, 592(7852): 86--92.

\bibitem[{Wang, Sporns, and Burkhalter(2012)}]{wang2012network}
Wang, Q.; Sporns, O.; and Burkhalter, A. 2012.
\newblock Network analysis of corticocortical connections reveals ventral and
  dorsal processing streams in mouse visual cortex.
\newblock \emph{Journal of Neuroscience}, 32(13): 4386--4399.

\bibitem[{Wightman(2019)}]{rw2019timm}
Wightman, R. 2019.
\newblock PyTorch Image Models.
\newblock \url{https://github.com/rwightman/pytorch-image-models}.
\newblock Accessed: 2022-01-17.

\bibitem[{Xie et~al.(2017)Xie, Girshick, Doll{\'a}r, Tu, and
  He}]{xie2017aggregated}
Xie, S.; Girshick, R.; Doll{\'a}r, P.; Tu, Z.; and He, K. 2017.
\newblock Aggregated residual transformations for deep neural networks.
\newblock In \emph{Proceedings of the IEEE Conference on Computer Vision and
  Pattern Recognition (CVPR)}, 1492--1500.

\bibitem[{Yamins and DiCarlo(2016)}]{yamins2016using}
Yamins, D.~L.; and DiCarlo, J.~J. 2016.
\newblock Using goal-driven deep learning models to understand sensory cortex.
\newblock \emph{Nature neuroscience}, 19(3): 356--365.

\bibitem[{Yamins et~al.(2014)Yamins, Hong, Cadieu, Solomon, Seibert, and
  DiCarlo}]{yamins2014performance}
Yamins, D.~L.; Hong, H.; Cadieu, C.~F.; Solomon, E.~A.; Seibert, D.; and
  DiCarlo, J.~J. 2014.
\newblock Performance-optimized hierarchical models predict neural responses in
  higher visual cortex.
\newblock \emph{Proceedings of the national academy of sciences}, 111(23):
  8619--8624.

\bibitem[{Zagoruyko and Komodakis(2016)}]{zagoruyko2016wide}
Zagoruyko, S.; and Komodakis, N. 2016.
\newblock Wide residual networks.
\newblock In \emph{Proceedings of the British Machine Vision Conference
  (BMVC)}.

\bibitem[{Zhu and Yang(2013)}]{zhu2013multi}
Zhu, X.; and Yang, Z. 2013.
\newblock Multi-scale spatial concatenations of local features in natural
  scenes and scene classification.
\newblock \emph{Plos one}, 8(9): e76393.

\bibitem[{Zhuang et~al.(2021)Zhuang, Yan, Nayebi, Schrimpf, Frank, DiCarlo, and
  Yamins}]{zhuang2021unsupervised}
Zhuang, C.; Yan, S.; Nayebi, A.; Schrimpf, M.; Frank, M.~C.; DiCarlo, J.~J.;
  and Yamins, D.~L. 2021.
\newblock Unsupervised neural network models of the ventral visual stream.
\newblock \emph{Proceedings of the National Academy of Sciences}, 118(3):
  e2014196118.

\bibitem[{Zoccolan et~al.(2009)Zoccolan, Oertelt, DiCarlo, and
  Cox}]{zoccolan2009rodent}
Zoccolan, D.; Oertelt, N.; DiCarlo, J.~J.; and Cox, D.~D. 2009.
\newblock A rodent model for the study of invariant visual object recognition.
\newblock \emph{Proceedings of the National Academy of Sciences}, 106(21):
  8748--8753.

\end{thebibliography}

\clearpage

\appendix

\setcounter{secnumdepth}{1}

\section{Implementation Details of SNNs}
\label{appendix.snn_train}

\subsection{Spiking Neuron Model}

For all SNNs, we use the Integrate-and-Fire (IF) model as the spiking neuron model, which acts as the activation layer in neural networks. As mentioned in \cite{fang2021incorporating, fang2021deep}, $V_t$, $X_t$ and $S_t$ denote the state (membrane voltage), input (current) and output (spike) of the spiking neuron model respectively at time-step $t$, and the dynamics of the IF model can be described as follows:

\begin{align}
	H_t &= V_{t - 1} + X_t, \\
	S_t &= \Theta(H_t - V_{thresh}), \\
	V_t &= H_t (1 - S_t) + V_{reset} S_t.
\end{align}

While $V_t$ is the membrane voltage after the trigger of a spike, $H_t$ is also the membrane voltage, but after charging and before a spike firing. $\Theta(x)$ is the unit step function, so $S_t$ equals 1 when $H_t$ is greater than or equal to the threshold voltage $V_{thresh}$ and 0 otherwise. Meanwhile, when a spike fires, $V_t$ is reset to $V_{reset}$. Here, we set $V_{thresh} = 1$ and $V_{reset} = 0$.

In addition, because $\Theta(x)$ is non-differentiable at 0, the surrogate gradient method \cite{neftci2019surrogate} is applied to approximate the derivative function during back-propagation. Here, we use the inverse tangent function as the surrogate gradient function 

\begin{align}
	\sigma(x) = \frac{1}{\pi}  \arctan(\pi x) + \frac{1}{2},
\end{align}

and the derivative function is 

\begin{align}
	{\sigma}^{\prime}(x) = \frac{1}{1 + (\pi x)^2}.
\end{align}

\subsection{Architectures of SNNs}

\begin{figure*}[t]
	\centering
	\subfigure{
		\includegraphics[width=0.99\textwidth]{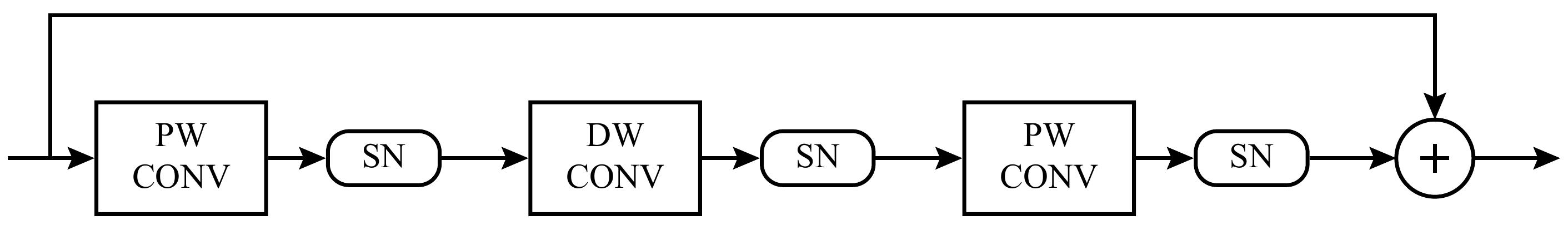}}
	\caption{The basic block of SpikingMobileNet. "PW CONV" is the pointwise convolution and "DW CONV" is the depthwise convolution. "SN" is the spiking neuron.}
	\label{Fig.spikingmobilenet}
\end{figure*}

In our experiments on SNNs, we not only use SEW ResNet proposed by \cite{fang2021deep}, but also build several new SNNs. On the one hand, we improve the spike-element-wise block in SEW ResNet with new architectures referring to studies on ResNet \cite{he2016deep, zagoruyko2016wide, xie2017aggregated}, as shown in Table \ref{table.snn}. On the other hand, as the multi-branch structures in CNNs increase neural representation similarity to mouse visual cortex, we use depthwise separable convolutions and follow the overall architecture of MobileNetV2 \cite{howard2017mobilenets, sandler2018mobilenetv2} to build the SpikingMobileNet, the basic block of which is shown in Figure \ref{Fig.spikingmobilenet}.

Our implementation is based on SpikingJelly \cite{SpikingJelly}, an open-source framework of deep SNN.

\subsection{Hyper-Parameters in SNNs' Training}

We use the ImageNet dataset to pre-train the new SNNs. Following the settings for training SEW ResNet \cite{fang2021deep}, we train the models for 320 epochs on 8 GPUs (NVIDIA V100), using SGD with a mini-batch size of 32. The momentum is 0.9 and the weight decay is 0. The initial learning rate is 0.1 and we decay it with a cosine annealing, where the maximum number of iterations is the same as the number of epochs. For all SNNs, we set the simulation duration $T = 4$.  

\section{Overall model rankings}
\label{appendix.overall_rankings}

\begin{figure*}[t]
	\centering
	\includegraphics[width=0.80\textwidth]{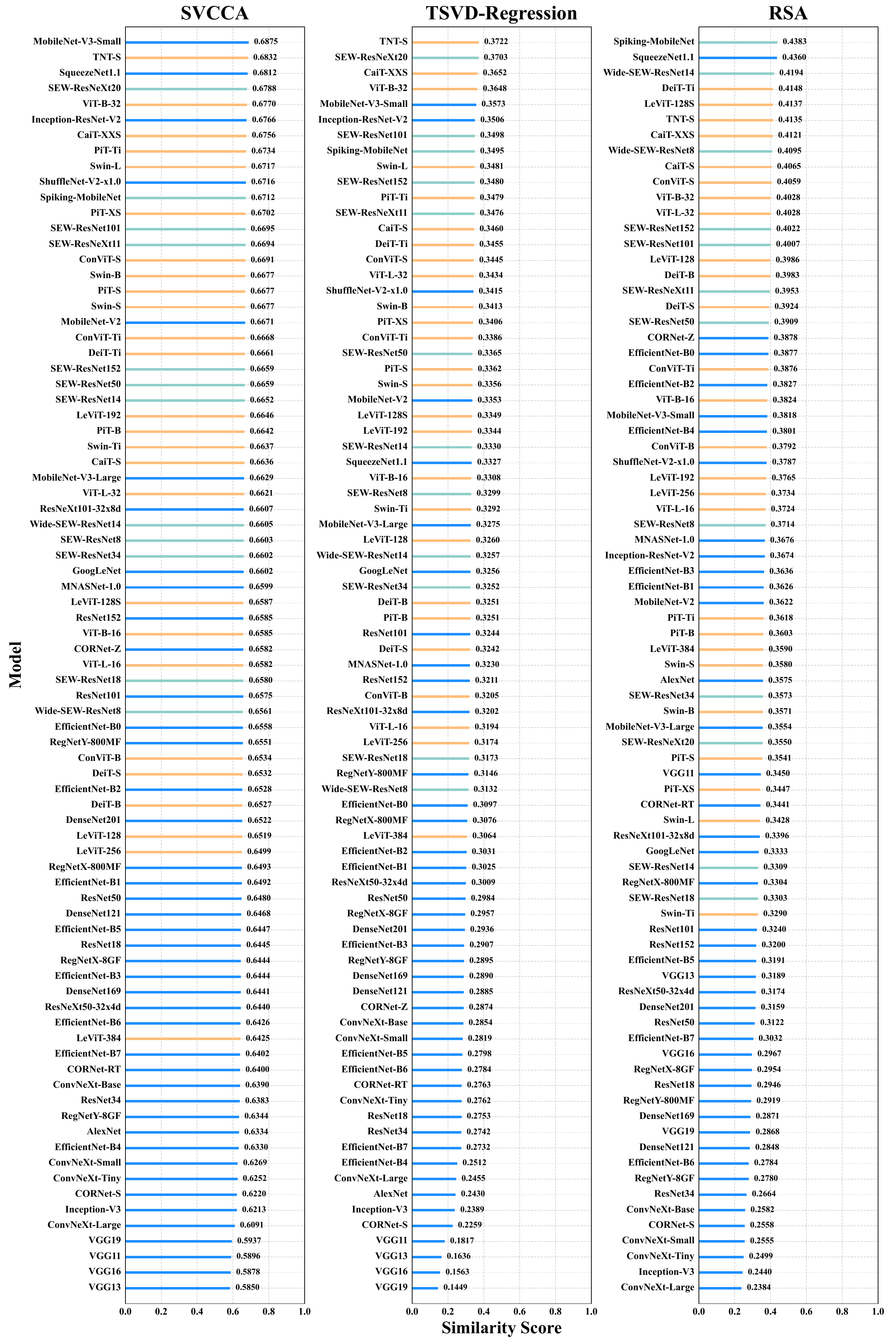}
	\caption{Overall model rankings of the similarity scores on Allen Brain mouse dataset. The similarity scores of CNNs, SNNs and vision transformers are shown by blue, green and orange bars, respectively.}
	\label{Fig.model_rank_allen_brain}
\end{figure*}

\begin{figure*}[t]
	\centering
	\includegraphics[width=0.80\textwidth]{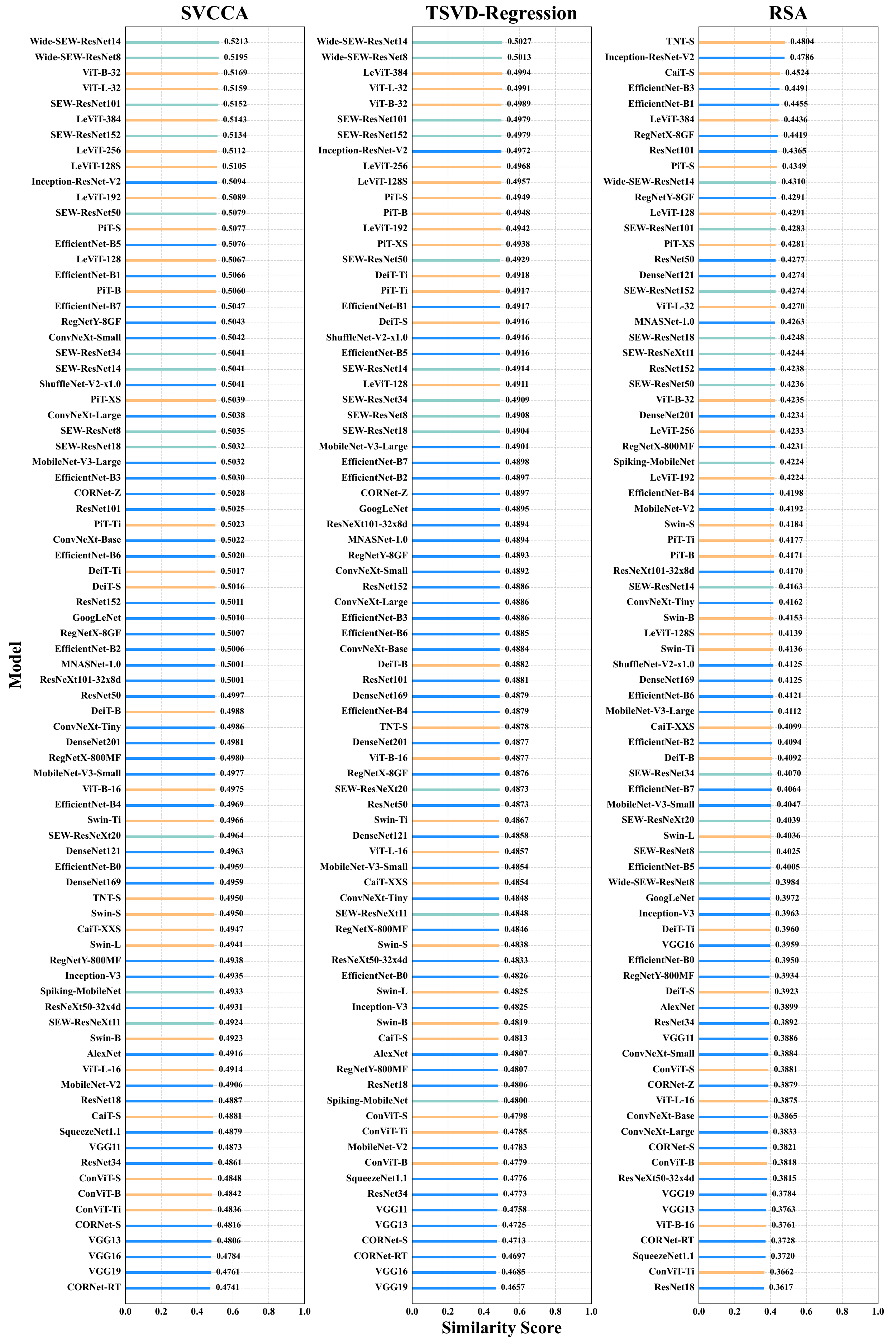}
	\caption{Overall model rankings of the similarity scores on Macaque-Face dataset.}
	\label{Fig.model_rank_macaque_face}
\end{figure*}

\begin{figure*}[t]
	\centering
	\includegraphics[width=0.80\textwidth]{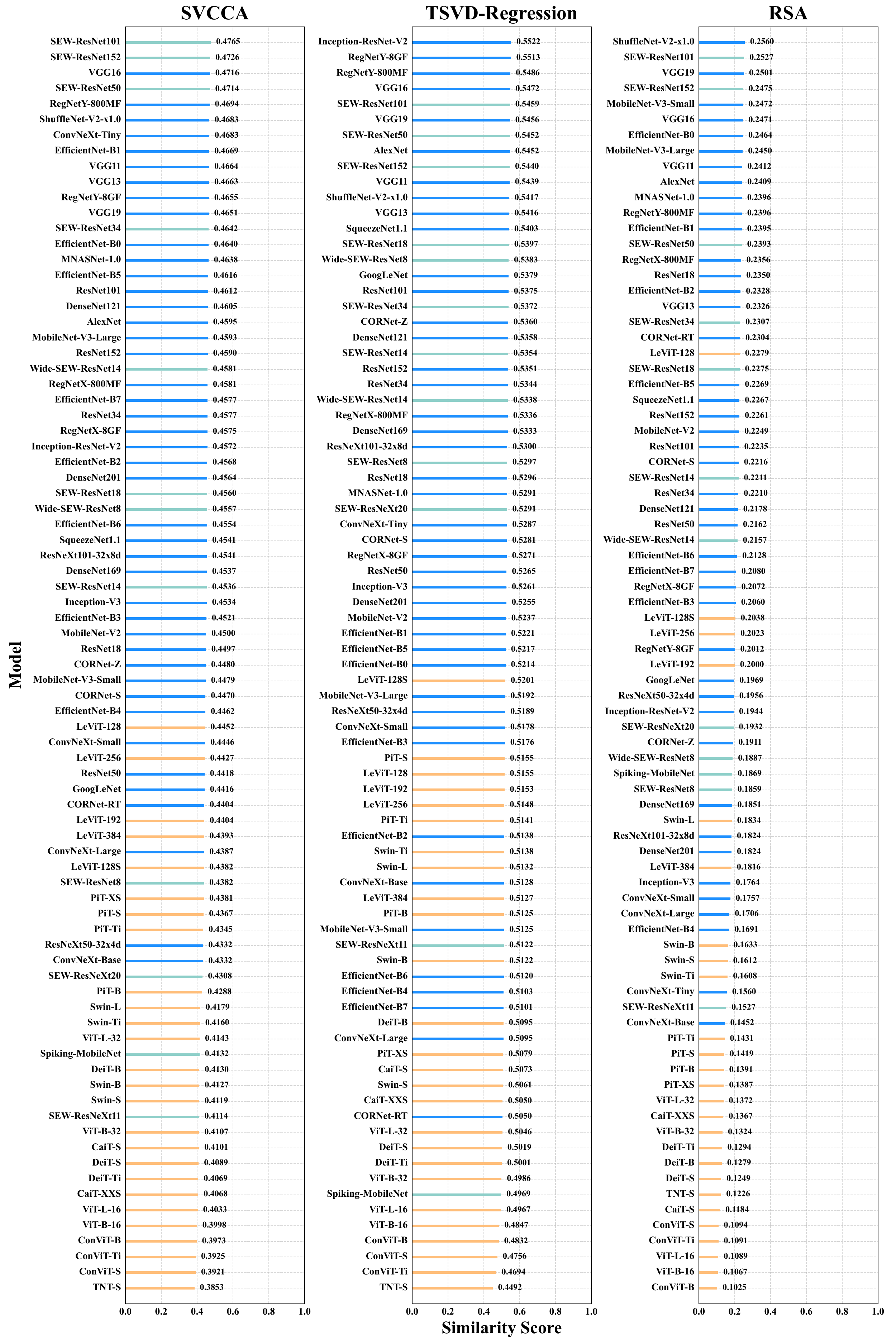}
	\caption{Overall model rankings of the similarity scores on Macaque-Synthetic dataset.}
	\label{Fig.model_rank_macaque_synthetic}
\end{figure*}

\begin{figure*}[t]
	\centering
	\includegraphics[width=0.99\textwidth]{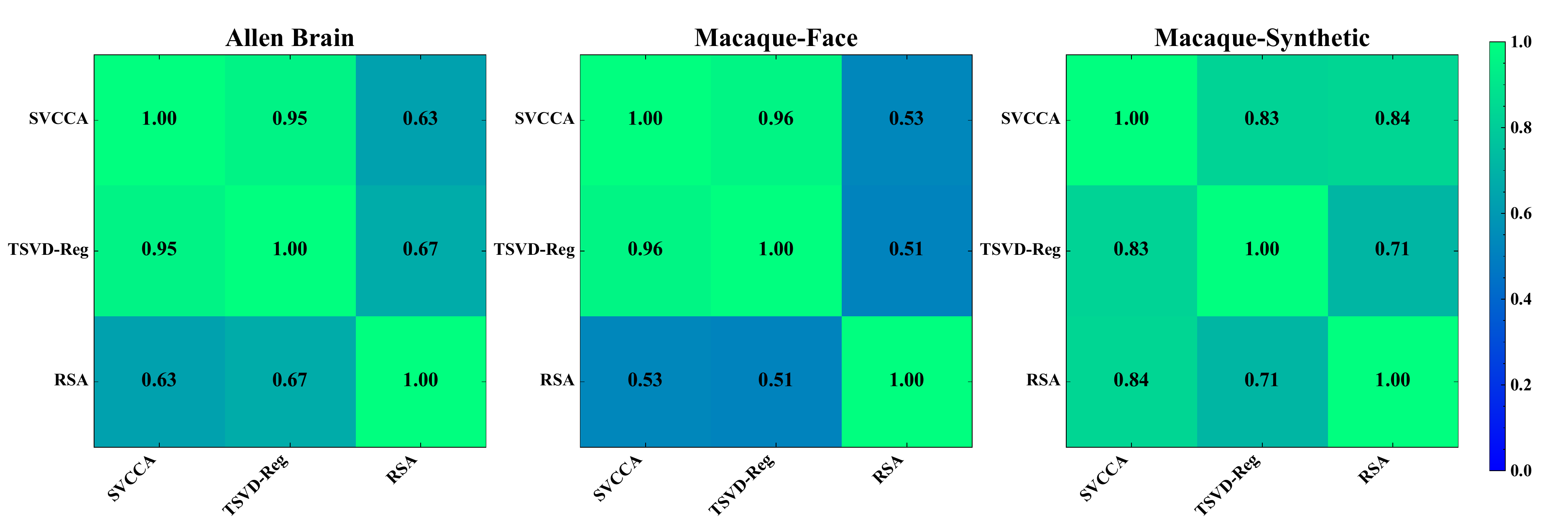}
	\caption{The Spearman's rank correlation between the overall model rankings of different metrics. There is a strong correlation between SVCCA and TSVD-Reg, but RSA has weaker correlations with them.}
	\label{Fig.rank_order}
\end{figure*}

The results of model rankings are shown in Figure \ref{Fig.model_rank_allen_brain}, \ref{Fig.model_rank_macaque_face} and \ref{Fig.model_rank_macaque_synthetic}. We also apply the Spearman's rank correlation to the overall model rankings of different metrics, which is shown in Figure \ref{Fig.rank_order}.

\section{Score Comparisons among Model Groups}
\label{appendix.model_type_compare}

\begin{figure*}[t]
	\centering
	\includegraphics[width=0.99\textwidth]{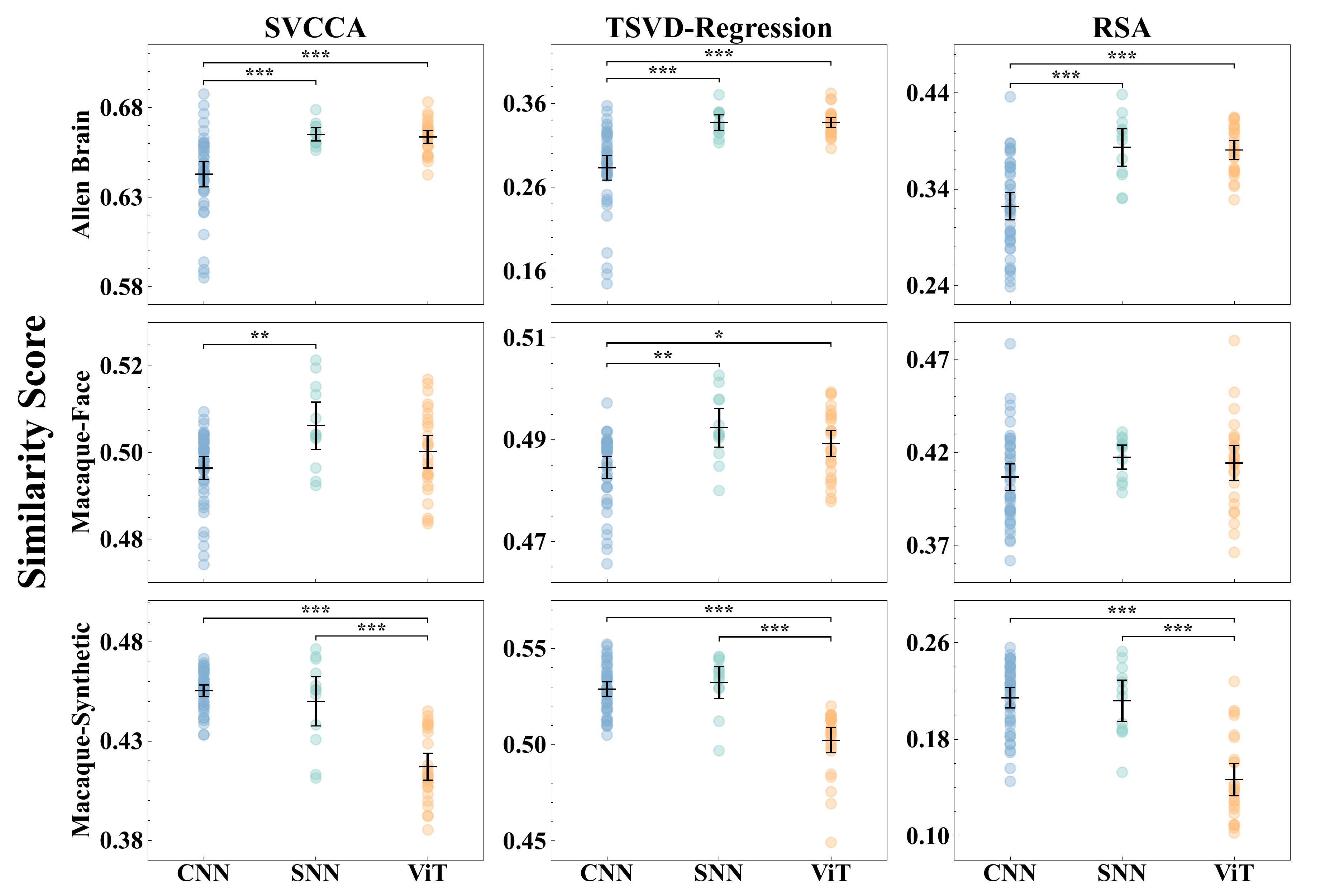}
	\caption{For three datasets and three similarity metrics, each point indicates the final representation similarity score of a model. There are three types of models (CNNs, SNNs, and vision transformers). From top to bottom, each row respectively corresponds to Allen Brain dataset, Macaque-Face dataset, and Macaque-Synthetic dataset. The error bar is the 95\% confidence interval across models from a group. The '*' on the horizontal brackets indicates how significant the differences are.}
	\label{Fig.model_type_compare}
\end{figure*}

We conduct comparisons of similarity scores among CNNs, SNNs, and vision transformers. The results are shown in Figure \ref{Fig.model_type_compare}.

\section{Overall CNN rankings}
\label{appendix.cnn_rankings}

\begin{figure*}[t]
	\centering
	\includegraphics[width=0.99\textwidth]{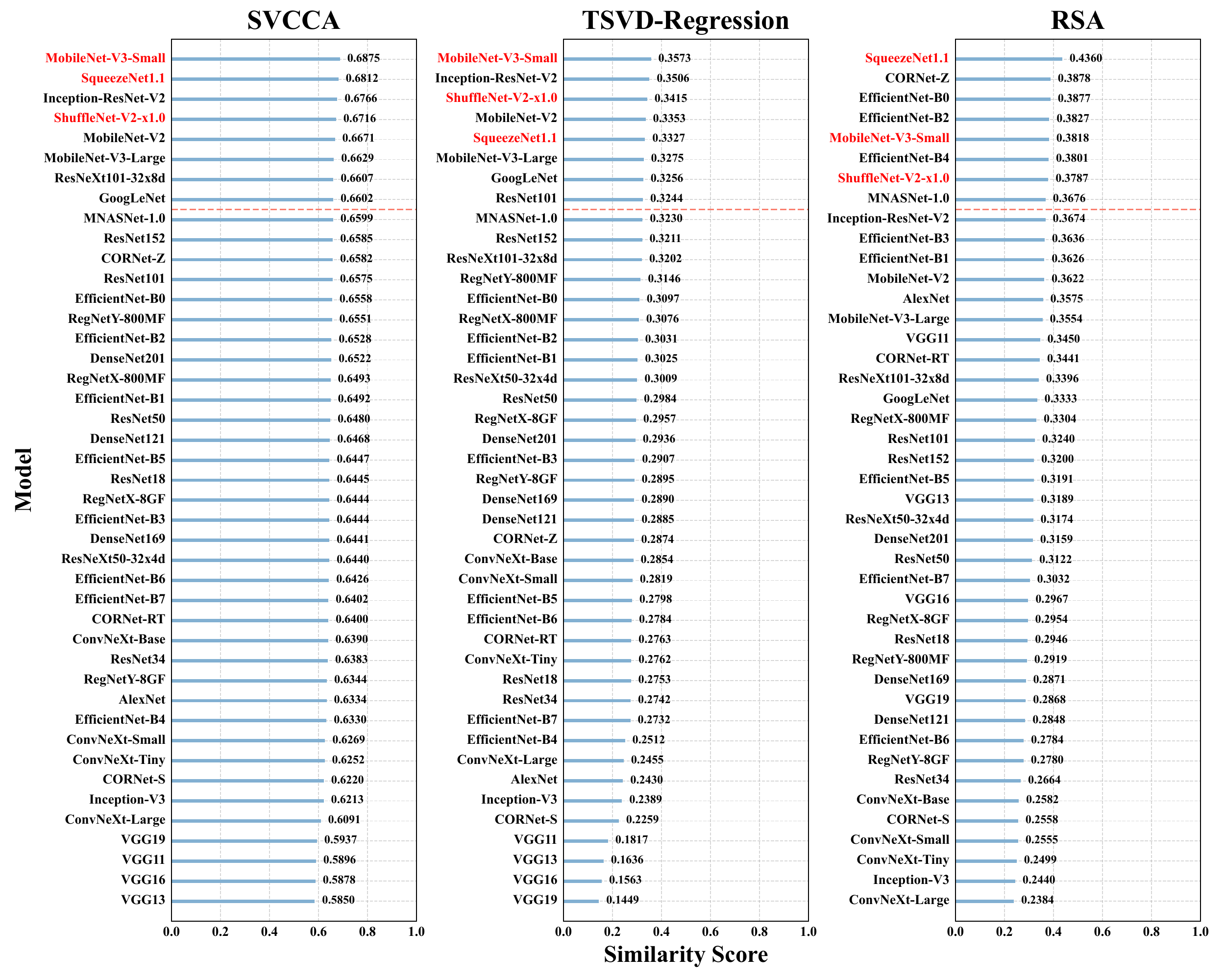}
	\caption{Overall CNN rankings of the similarity scores on Allen Brain mouse dataset. The top 20\% models are above the red dotted lines. The networks that are in the top 20\% for all three metrics are marked in red.}
	\label{Fig.cnn_model_rank_allen_brain}
\end{figure*}

\begin{figure*}[t]
	\centering
	\subfigure{
		\includegraphics[width=0.99\textwidth]{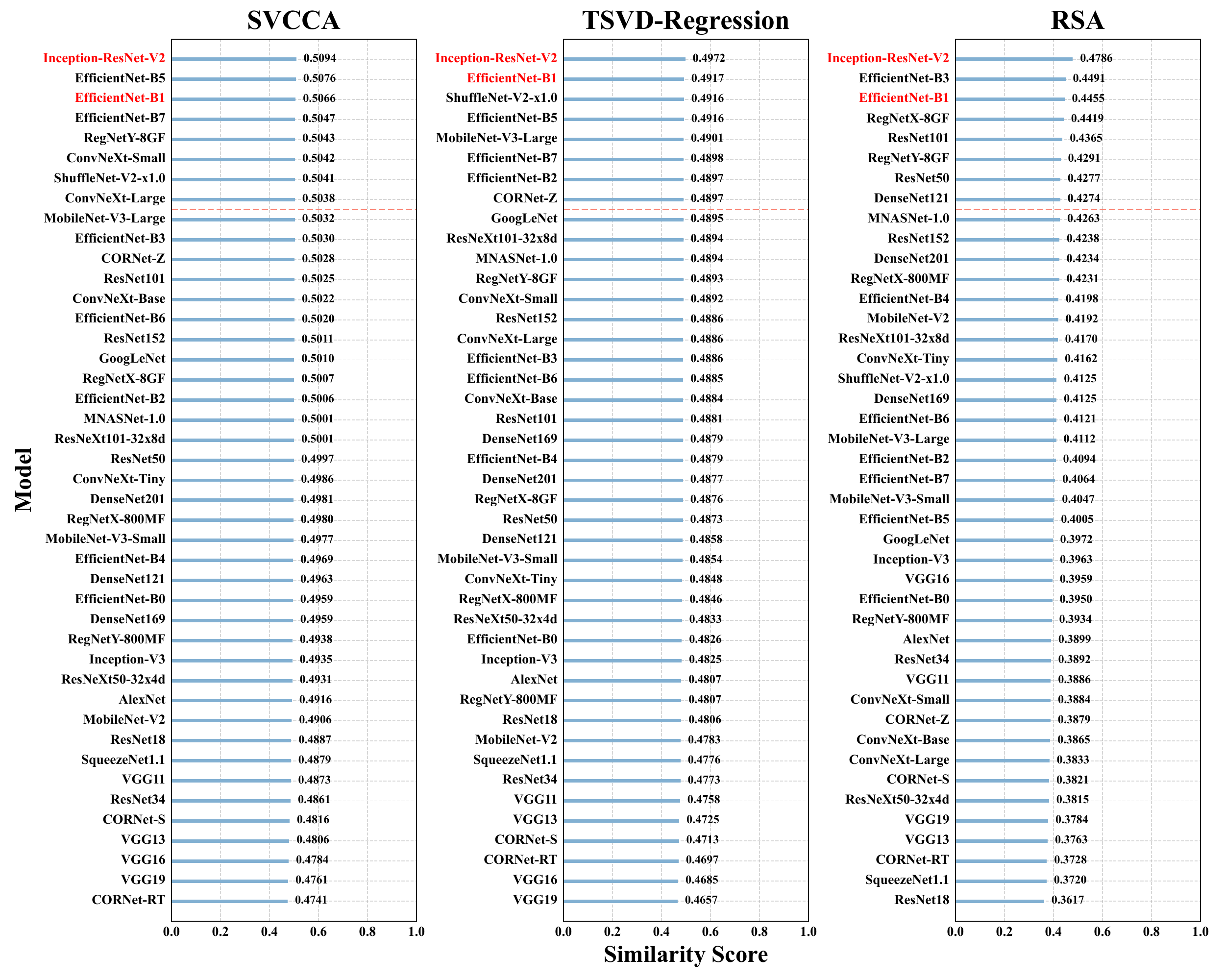}}
	\caption{Overall CNN rankings of the similarity scores on Macaque-Face dataset.}
	\label{Fig.cnn_model_rank_macaque_face}
\end{figure*}

\begin{figure*}[t]
	\centering
	\includegraphics[width=0.99\textwidth]{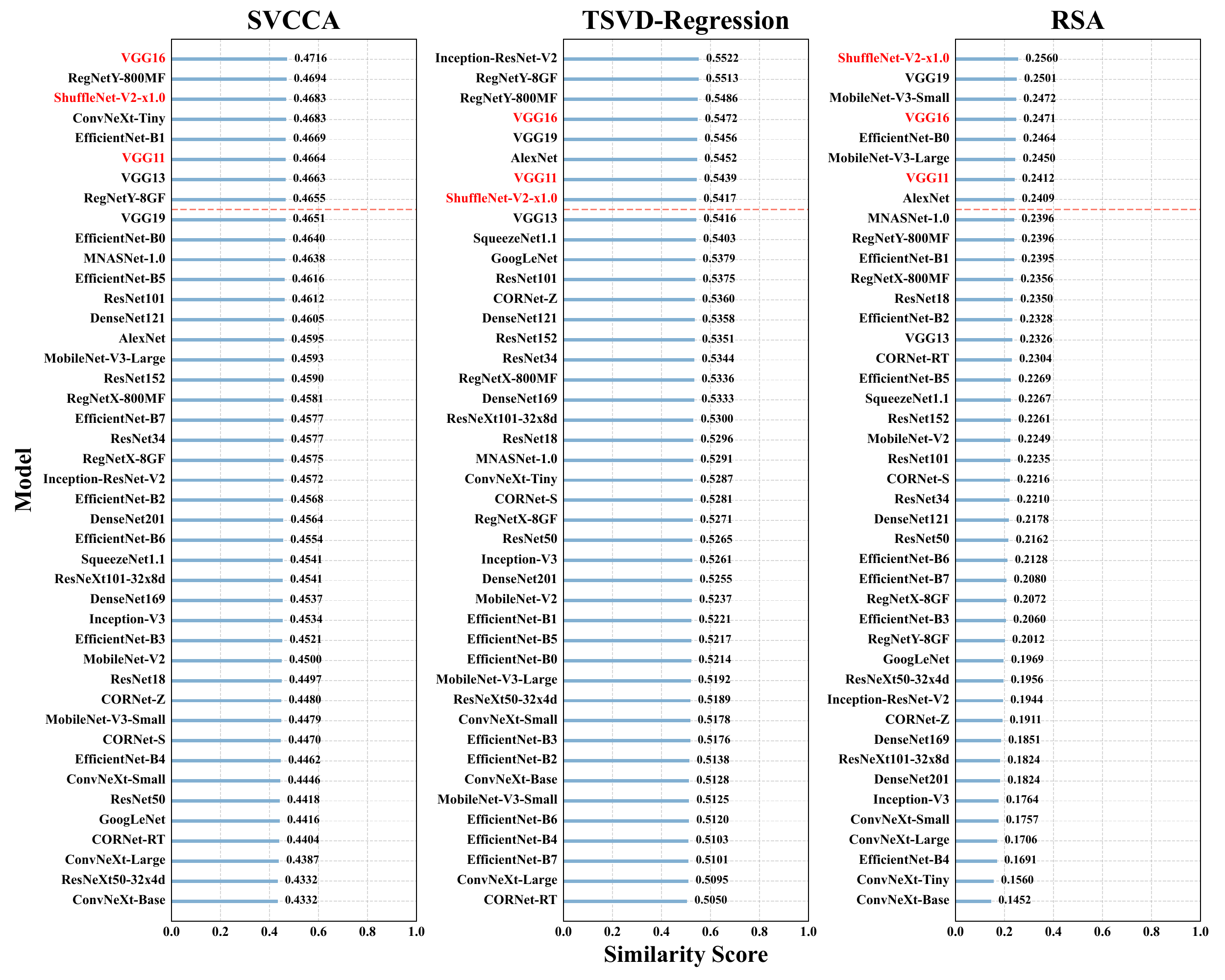}
	\caption{Overall CNN rankings of the similarity scores on Macaque-Synthetic dataset.}
	\label{Fig.cnn_model_rank_macaque_synthetic}
\end{figure*}

The results of CNN rankings are shown in Figure \ref{Fig.cnn_model_rank_allen_brain}, \ref{Fig.cnn_model_rank_macaque_face} and \ref{Fig.cnn_model_rank_macaque_synthetic}.

\section{Correlations between the Model Sizes and the Similarity Scores}
\label{appendix.model_size_correlation}

\begin{figure*}[t]
	\centering
	\subfigure{
		\includegraphics[width=0.49\textwidth]{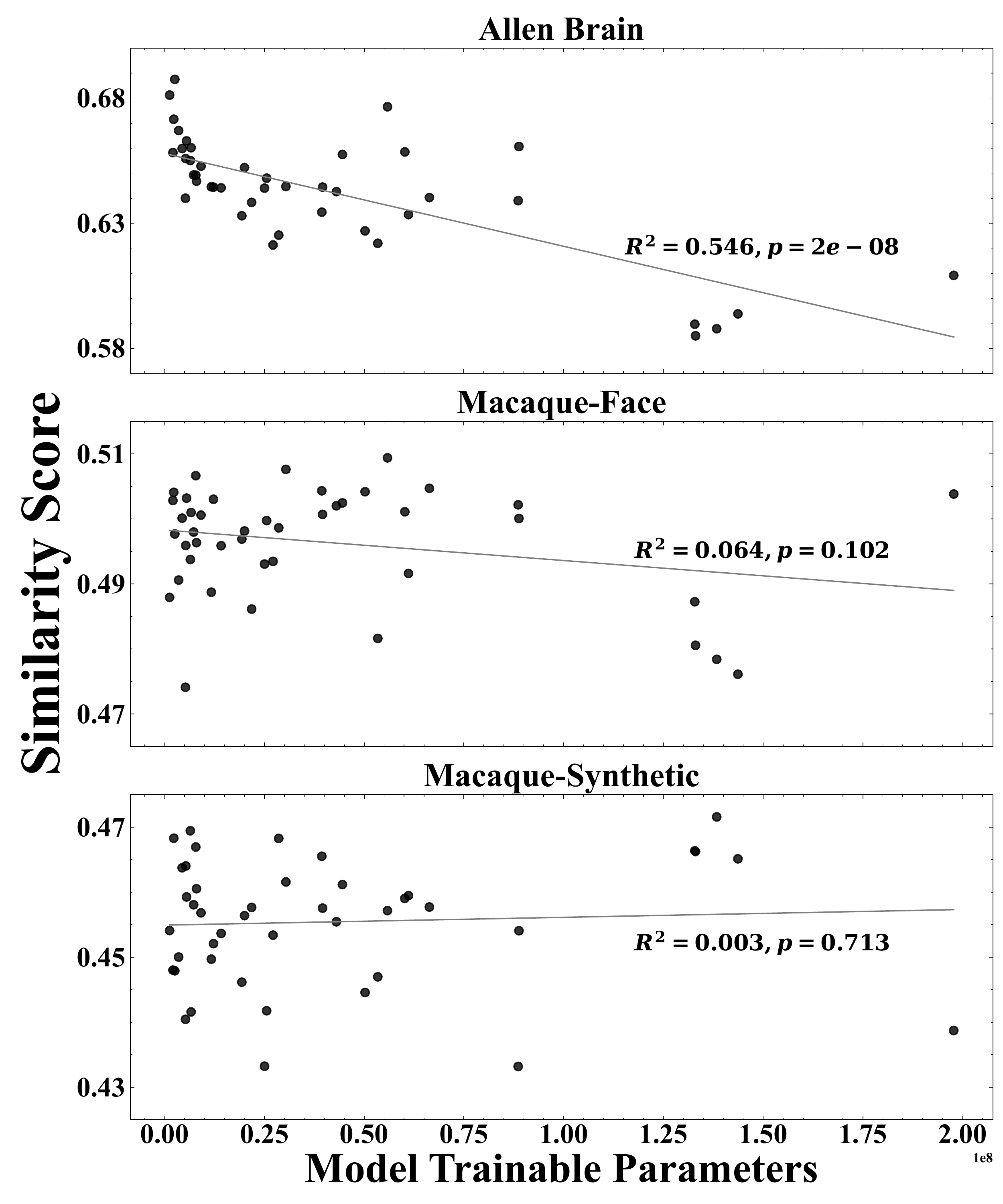}}
	\subfigure{
		\includegraphics[width=0.49\textwidth]{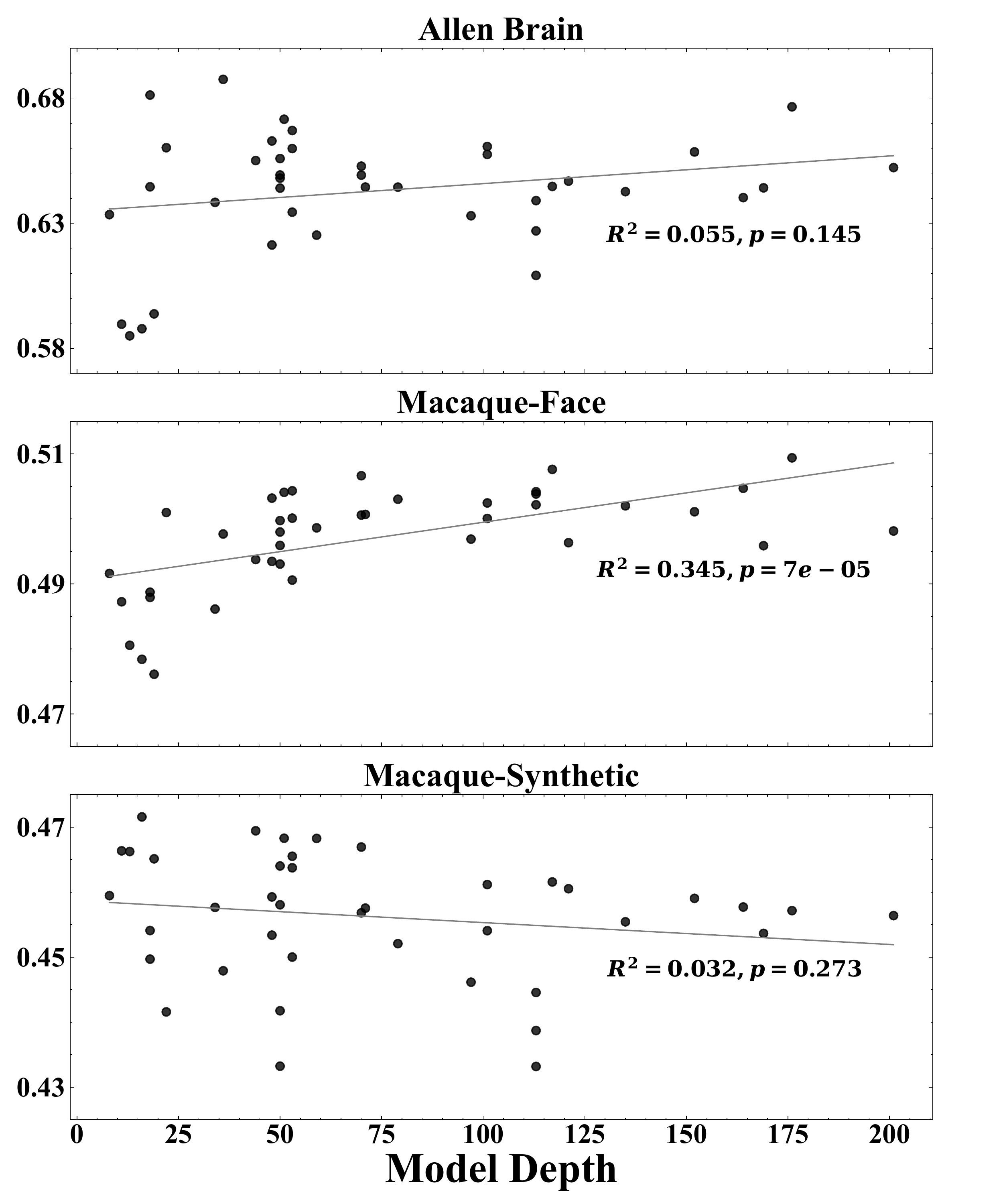}}
	\caption{The results of linear regression to model sizes and the similarity scores for SVCCA. Each point indicates a model.}
	\label{Fig.model_size_compare_svcca}
\end{figure*}

\begin{figure*}[t]
	\centering
	\subfigure{
		\includegraphics[width=0.49\textwidth]{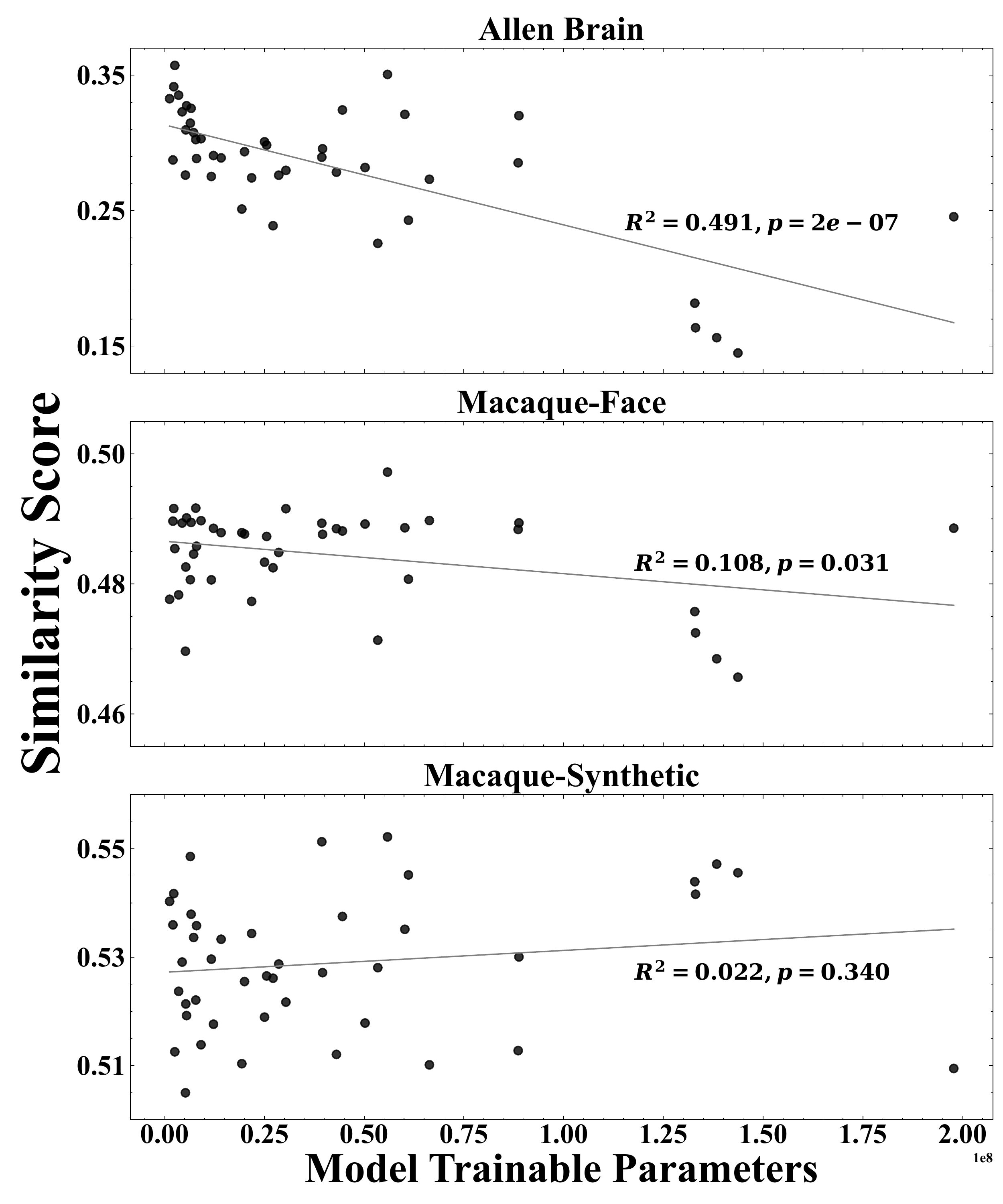}}
	\subfigure{
		\includegraphics[width=0.49\textwidth]{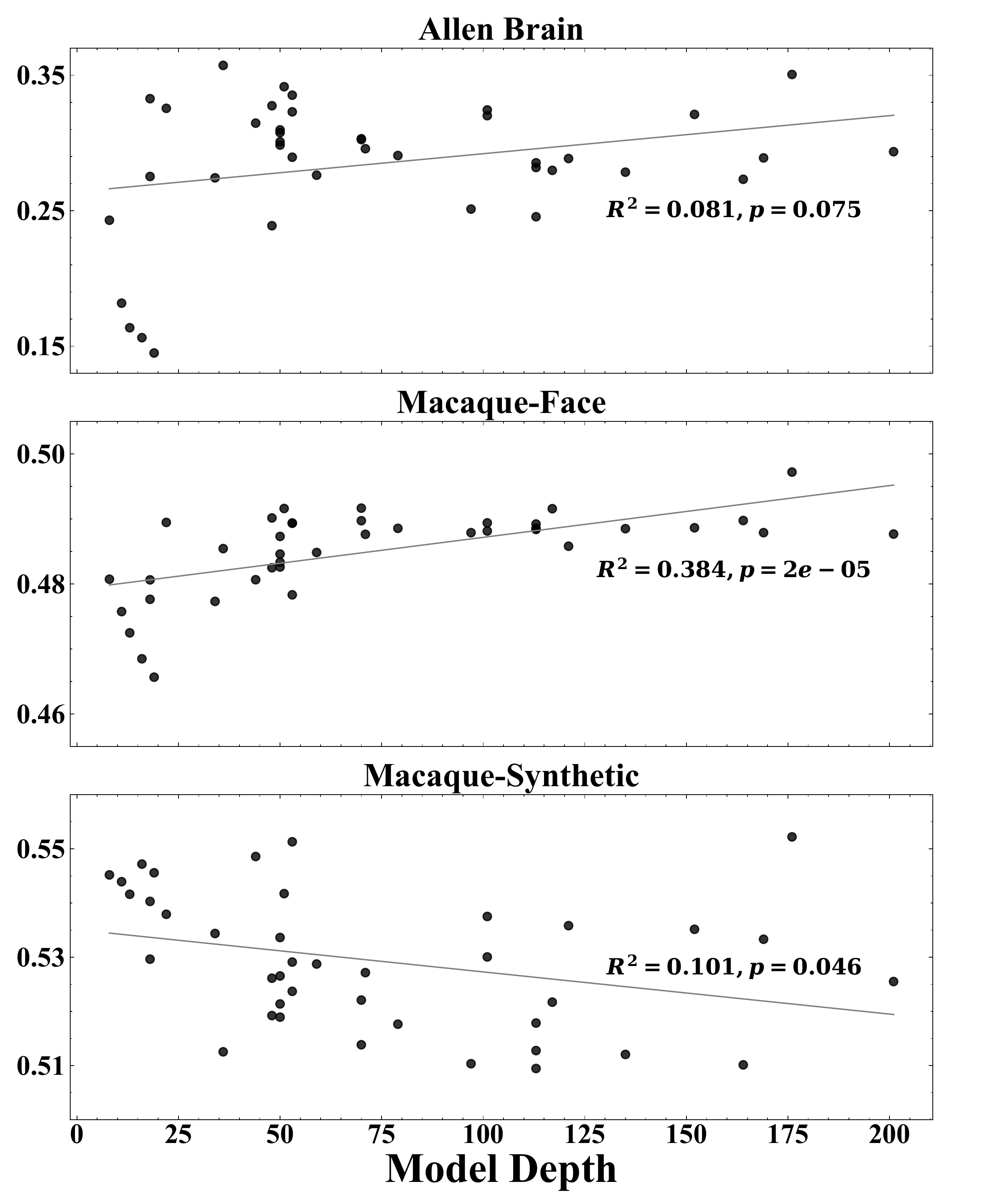}}
	\caption{The results of linear regression to model sizes and the similarity scores for TSVD-Reg.}
	\label{Fig.model_size_compare_reg}
\end{figure*}

\begin{figure*}[t]
	\centering
	\subfigure{
		\includegraphics[width=0.49\textwidth]{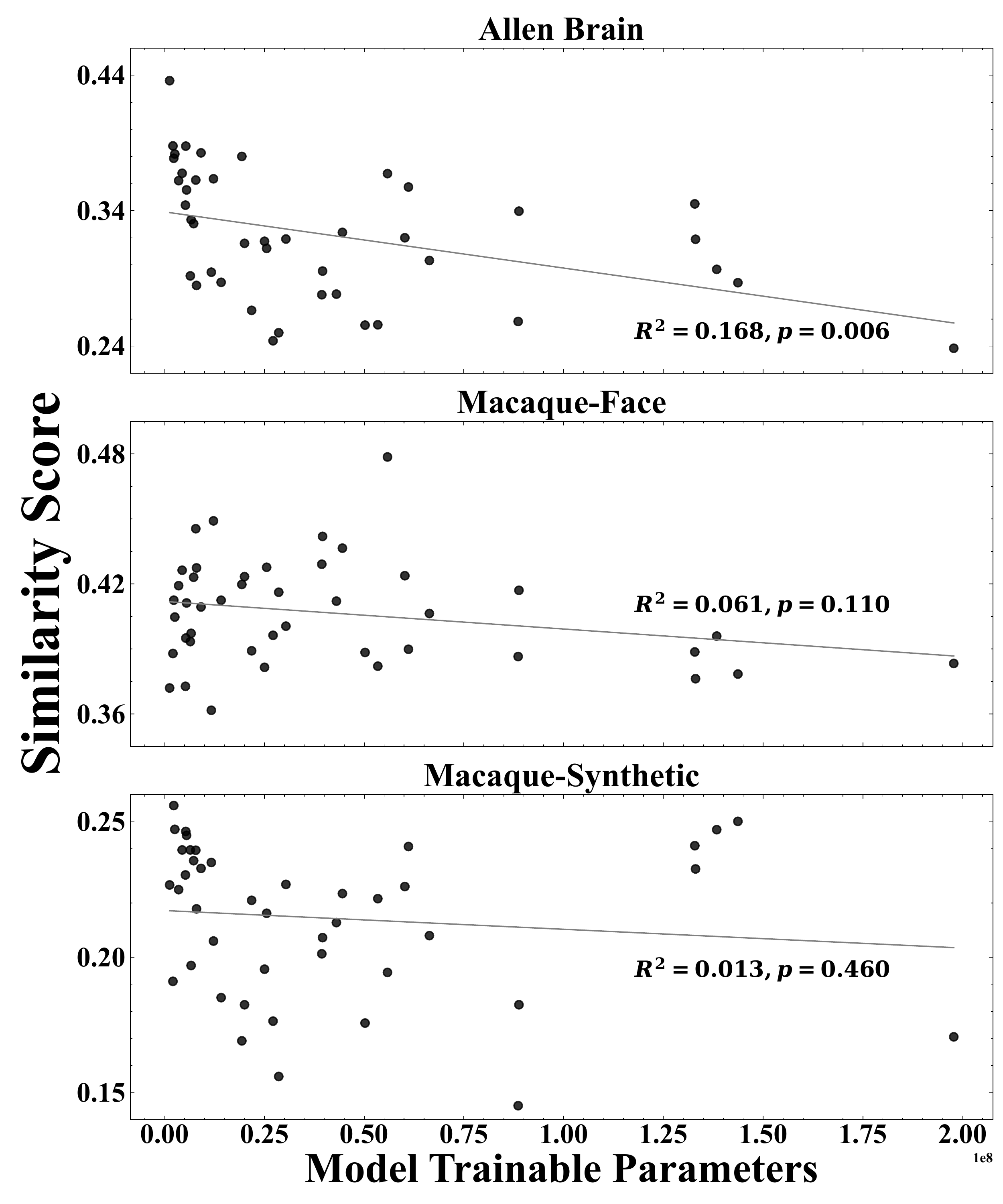}}
	\subfigure{
		\includegraphics[width=0.49\textwidth]{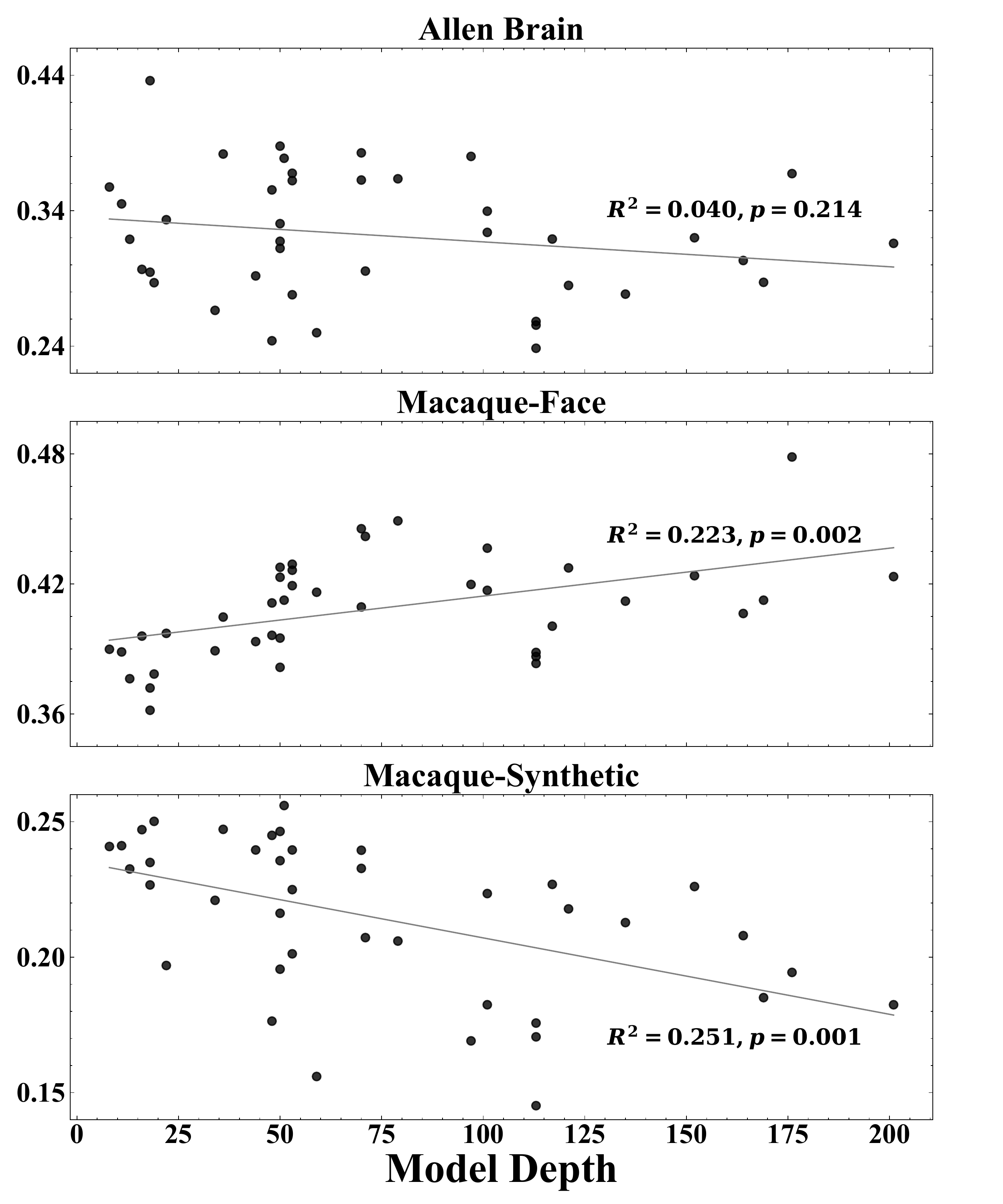}}
	\caption{The results of linear regression to model sizes and the similarity scores for RSA.}
	\label{Fig.model_size_compare_rsa}
\end{figure*}

The results of linear regression to model sizes and the similarity scores are shown in Figure \ref{Fig.model_size_compare_svcca}, \ref{Fig.model_size_compare_reg} and \ref{Fig.model_size_compare_rsa}.

\section{The ImageNet Accuracy and the Similarity Scores}
\label{appendix.ImageNet accuracy}

\begin{figure*}[t]
	\centering
	\subfigure[SVCCA]{
		\includegraphics[width=0.32\textwidth]{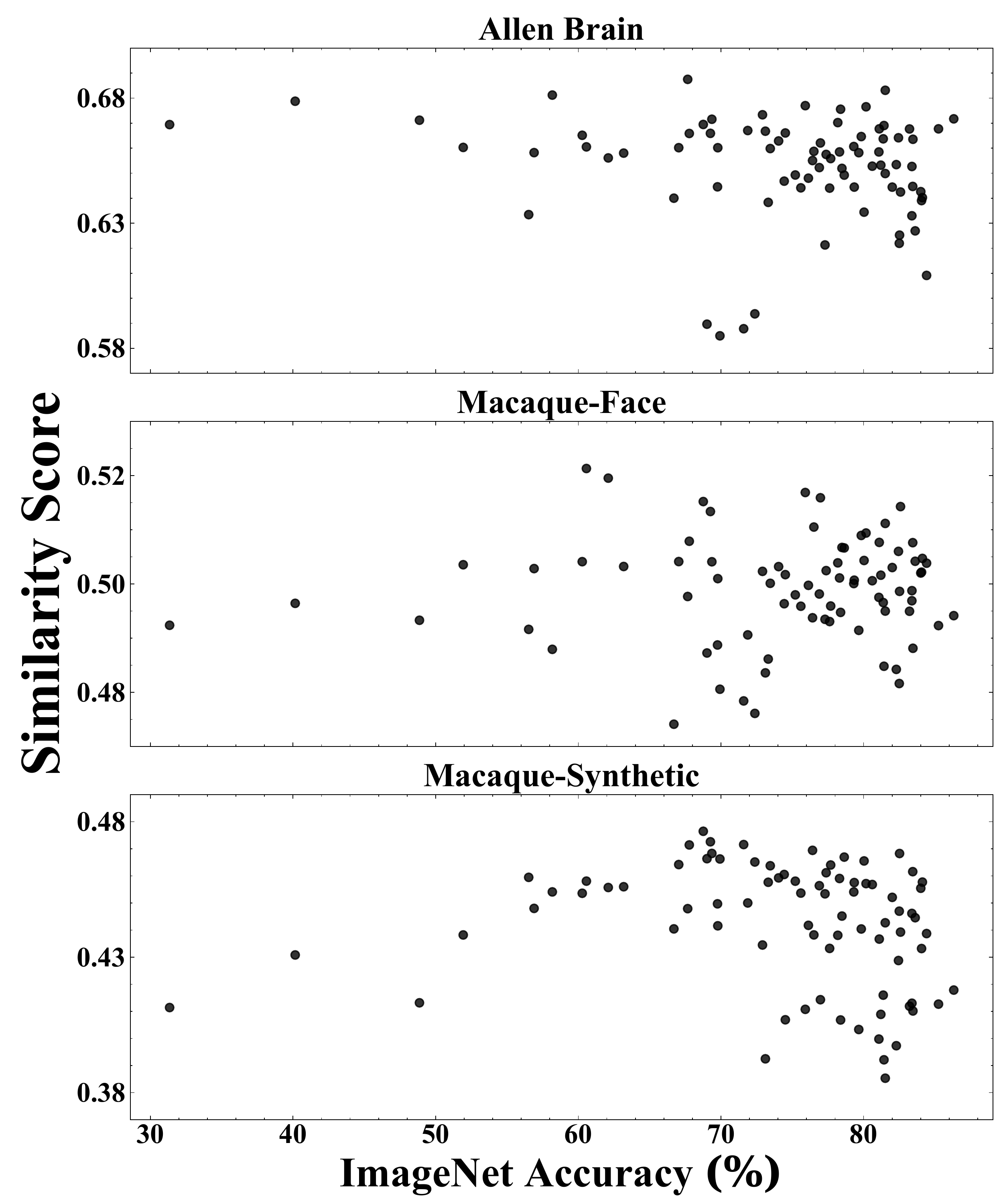}}
	\subfigure[TSVD-Reg]{
		\includegraphics[width=0.32\textwidth]{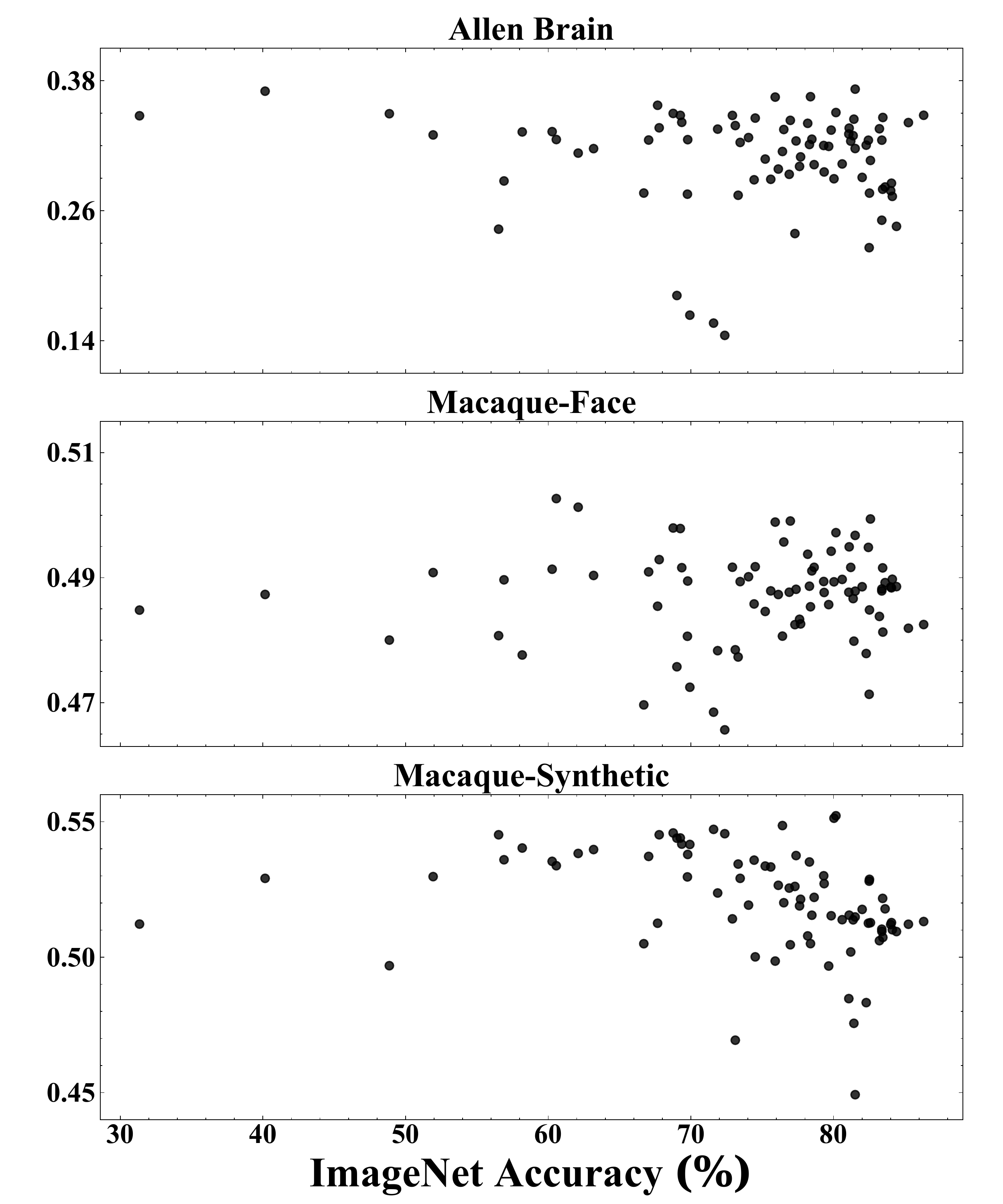}}
	\subfigure[RSA]{
		\includegraphics[width=0.32\textwidth]{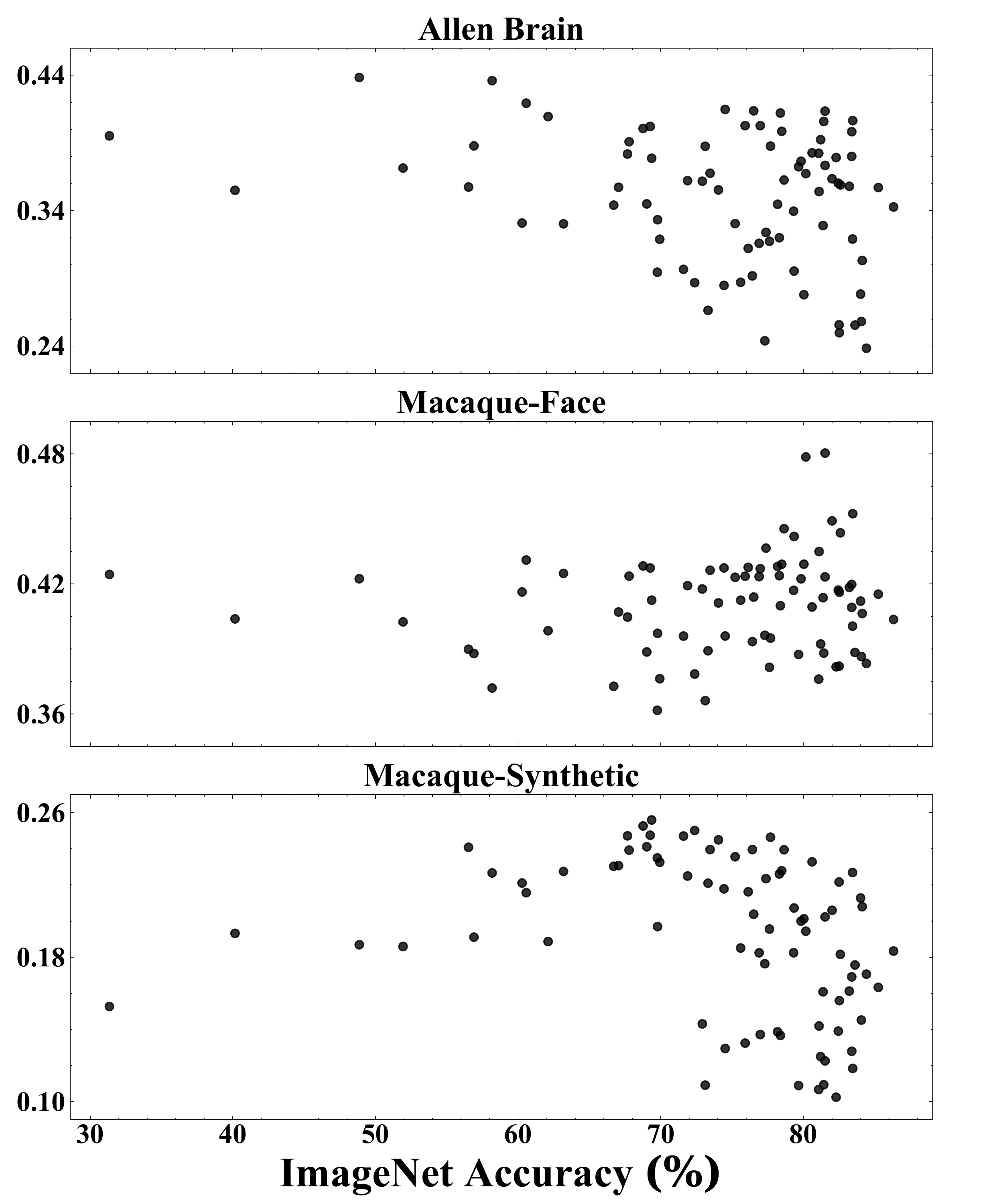}}
	\caption{The ImageNet accuracy and the similarity scores for SVCCA, TSVD-Reg and RSA.}
	\label{Fig.imagenet_accuracy}
\end{figure*}

The results are shown in Figure \ref{Fig.imagenet_accuracy}.

\begin{table*}[t]
	\centering
	\scalebox{0.7}
	{
		\begin{tabular}{|c|c|c|c|c|c|c|c|}
			\hline
			layers & output size & SEW ResNet8 & SEW ResNet14 & Wide SEW ResNet8 & Wide SEW ResNet14 & SEW ResNeXt11 & SEW ResNeXt20 \\
			\hline
			conv with bn & $112 \times 112$ & \multicolumn{6}{c|}{$7 \times 7$, 64, stride 2} \\
			\hline
			sn & $112 \times 112$ & \multicolumn{6}{c|}{} \\
			\hline
			max pool & $56 \times 56$ & \multicolumn{6}{c|}{$3 \times 3$, stride 2} \\
			\hline
			SEW block1 & $28 \times 28$ & 
			$\begin{bmatrix}
				3 \times 3, 128 \\
				3 \times 3, 128
			\end{bmatrix} \times 1$ & 
			$\begin{bmatrix}
				3 \times 3, 128 \\
				3 \times 3, 128
			\end{bmatrix} \times 2$ & 
			$\begin{bmatrix}
				3 \times 3, 128 \times 4 \\
				3 \times 3, 128 \times 4
			\end{bmatrix} \times 1$ & 
			$\begin{bmatrix}
				3 \times 3, 128 \times 4 \\
				3 \times 3, 128 \times 4
			\end{bmatrix} \times 2$ & 
			$\begin{bmatrix}
				1 \times 1, 64 \\
				3 \times 3, 64, g = 32 \\
				1 \times 1, 128 \\
			\end{bmatrix} \times 1$ & 
			$\begin{bmatrix}
				1 \times 1, 64 \\
				3 \times 3, 64, g = 32 \\
				1 \times 1, 128 \\
			\end{bmatrix} \times 2$ \\
			\hline
			SEW block2 & $14 \times 14$ & 
			$\begin{bmatrix}
				3 \times 3, 256 \\
				3 \times 3, 256
			\end{bmatrix} \times 1$ & 
			$\begin{bmatrix}
				3 \times 3, 256 \\
				3 \times 3, 256
			\end{bmatrix} \times 2$ & 
			$\begin{bmatrix}
				3 \times 3, 256 \times 4 \\
				3 \times 3, 256 \times 4
			\end{bmatrix} \times 1$ & 
			$\begin{bmatrix}
				3 \times 3, 256 \times 4 \\
				3 \times 3, 256 \times 4
			\end{bmatrix} \times 2$ & 
			$\begin{bmatrix}
				1 \times 1, 128 \\
				3 \times 3, 128, g = 32 \\
				1 \times 1, 256 \\
			\end{bmatrix} \times 1$ & 
			$\begin{bmatrix}
				1 \times 1, 128 \\
				3 \times 3, 128, g = 32 \\
				1 \times 1, 256 \\
			\end{bmatrix} \times 2$ \\
			\hline
			SEW block3 & $7 \times 7$ & 
			$\begin{bmatrix}
				3 \times 3, 512 \\
				3 \times 3, 512
			\end{bmatrix} \times 1$ & 
			$\begin{bmatrix}
				3 \times 3, 512 \\
				3 \times 3, 512
			\end{bmatrix} \times 2$ & 
			$\begin{bmatrix}
				3 \times 3, 512 \times 4 \\
				3 \times 3, 512 \times 4
			\end{bmatrix} \times 1$ & 
			$\begin{bmatrix}
				3 \times 3, 512 \times 4 \\
				3 \times 3, 512 \times 4
			\end{bmatrix} \times 2$ & 
			$\begin{bmatrix}
				1 \times 1, 256 \\
				3 \times 3, 256, g = 32 \\
				1 \times 1, 512 \\
			\end{bmatrix} \times 1$ & 
			$\begin{bmatrix}
				1 \times 1, 256 \\
				3 \times 3, 256, g = 32 \\
				1 \times 1, 512 \\
			\end{bmatrix} \times 2$ \\
			\hline
			\thead{global average \\ pool} & $1 \times 1$ & \multicolumn{6}{c|}{$7 \times 7$} \\
			\hline
			fc & $1000$ & \multicolumn{6}{c|}{} \\
			\hline
		\end{tabular}
	}
	\caption{Architectures of SNNs. "sn" denotes the spiking neuron. "$g = 32$" denotes the grouped convolutions with 32 groups. The hyper-parameters of the spike-element-wise block are shown in the brackets with the number of stacked blocks outside.}
	\label{table.snn}
\end{table*}

\end{document}